\documentclass[useAMS,usenatbib]{mn2e}
\usepackage{epsfig}
\usepackage[usenames,dvips]{color}
\usepackage{amsmath}
\usepackage{graphicx}
\usepackage{subfigure}
\usepackage{amssymb}
\usepackage{aas_macros}

\usepackage{multirow}
\usepackage{color}

\newcommand{\Msolar}{\mbox{\,$\rm M_{\odot}$}} 

\def\simge{\mathrel{\raise1.16pt\hbox{$>$}\kern-7.0pt \lower3.06pt\hbox{{$\scriptstyle\sim$}}}} 
\def\simle{\mathrel{\raise1.16pt\hbox{$<$}\kern-7.0pt \lower3.06pt\hbox{{$\scriptstyle \sim$}}}} 

\begin{document}

\title[Gravitational waves from double neutron stars]{The gravitational-wave signal generated by a galactic population of double neutron-star binaries}
\author[]{Shenghua Yu$^{1,2}$\thanks{shenghuayu@bao.ac.cn}, C. Simon Jeffery$^{3,4}$\thanks{csj@arm.ac.uk}\\
$^1$National Astronomical Observatories, Chinese Academy of Sciences, Beijing 100012, China \\
$^2$The Key Laboratory of Radio Astronomy, Chinese Academy of Sciences \\
$^3$Armagh Observatory, College Hill, Armagh BT61 9DG, N. Ireland\\
$^4$School of Physics, Trinity College Dublin, Dublin 2, Ireland}

\date{Accepted . Received ; in original form }

\pagerange{\pageref{firstpage}--\pageref{lastpage}} \pubyear{2011}

\maketitle

\label{firstpage}

\begin{abstract}
We investigate the gravitational wave (GW) signal generated
by a population of double neutron-star binaries (DNS) with eccentric orbits
caused by kicks during supernova collapse and binary evolution.
The DNS population of a standard Milky-Way type galaxy has been studied
as a function of star formation history, initial mass function (IMF) and metallicity
and of the binary-star common-envelope ejection process.
The model provides birth rates, merger rates and total numbers of DNS as a function of time.
The GW signal produced by this population has been computed and expressed
in terms of a hypothetical space GW detector (eLISA)
by calculating the number of discrete GW signals at different confidence levels,
where `signal' refers to detectable GW strain in a given frequency-resolution element.
In terms of the parameter space explored, the number of DNS-originating
GW signals is greatest in regions of recent star formation, and is significantly increased
if metallicity is reduced from 0.02 to 0.001, consistent with \citet{Belczynski10a}.
Increasing the IMF power-law index (from --2.5 to --1.5)
increases the number of GW signals by a large factor.
This number is also much higher for
models where the common-envelope ejection is treated using the
$\alpha-$mechanism (energy conservation) than when using the  $\gamma-$mechanism
(angular-momentum conservation). We have estimated the total number of
detectable DNS GW signals from the Galaxy by combining contributions from thin disc, thick disc, bulge and halo.
The most probable numbers for an eLISA-type experiment are
 0$-$1600 signals per year at S/N$\geqslant$1, 0$-$900 signals per year at S/N$\geqslant$3, and 0$-$570
at S/N$\geqslant$5, coming from about 0$-$65, 0$-$60 and 0$-$50 resolved DNS respectively.
\end{abstract}


\section{Introduction}
\label{intro}

Most stars are members of binary or multiple star systems.
Over 70\% of massive stars (O-type stars) have a nearby companion
 which will affect their evolution, with over one half doing so before they leave the main sequence (MS)
\citep{sana12,Langer12}. If both members of a binary are massive enough to
end their evolution as core-collapse supernovae\footnote{Neutron stars can also be formed
by accretion of Oxygen-Neon white dwarfs}, after one or two explosions, the
final product could become a double neutron-star binary (DNS), a neutron-star
plus black-hole binary, or a double black-hole binary.
Evidence that such systems do form is provided by the double
pulsar PSR J0737-3039A/B \citep{Burgay03,Lyne04,Kramer08}.
Such compact binaries are expected to be a significant source of
gravitational wave (GW) radiation \citep{Barish99,Ricci97,Postnov06}.
They are amongst the sources most likely to be detected by a
gravitational wave detector in the frequency range $10^{-5} - 100$ Hz.
Aasi et al. (2014) measured upper limits of the GW strain
amplitudes from hundreds of pulsars using data from recent runs of the ground-based
GW observatories - LIGO, Virgo and GEO600, and showed that there are good prospects
for detections in the 10 - 1000 Hz range with the advanced LIGO and Virgo detectors.

Double neutron-star and black-hole binaries play an
important role in testing the theory of General Relativity, whilst double
pulsars  provide probes of magnetospheric physics.
The accurate timing of pulsars orbiting a
black hole  can be used to constrain the strain amplitude of
gravitational waves and the physical properties of the black hole.
\citep{Kramer04}.

\begin{table}
\caption{Orbital periods ($P_{\rm orb}$) and frequencies ($f_{\rm orb}$), eccentricities ($e$),  GW timescales ($\tau_{\rm GW}$)
and whether masses have been established for known and suspected DNS binaries;  after \citet{Lorimer08}. }
\label{tab_dns}
\begin{center}
\begin{tabular}{lrrcrl}
\hline
DNS & $P_{\rm orb}$ & $e$ & $M$  & $\log_{10} $ & $f_{\rm orb}$ \\
 \,\,PSR   &   (d)  &     &     &  $\tau_{\rm GW}{\rm yr^{-1}}$  & Hz \\
\hline
J0737--3039 & 0.102 & 0.09  & Yes & 7.9 & $1.1\times10^{-4}$ \\
J1906+0746 & 0.17 & 0.09 & Yes & 8.5 & $6.8\times10^{-5}$\\
B1913+16 & 0.3 & 0.62 & Yes & 8.5 &  $3.9\times10^{-5}$\\
B2127+11C & 0.3 & 0.68 & Yes & 8.3 &  $3.9\times10^{-5}$\\
J1756--2251 & 0.32 & 0.18 & Yes & 10.2 &  $3.6\times10^{-5}$\\
B1534+12 & 0.4 & 0.27 & Yes & 9.4 &  $2.9\times10^{-5}$\\
J1829+2456 & 1.18 & 0.14 & No & 10.8 &  $9.8\times10^{-6}$\\
J1518+4904 & 8.6 & 0.25 & No & 12.4 & $1.3\times10^{-6}$\\
J1811--1736 & 18.8 & 0.83 & Yes & 13.0 & $6.2\times10^{-7}$\\
B1820--11 & 357.8 & 0.79 & No & 15.8 & $3.2\times10^{-8}$\\
\hline
\end{tabular}
\\
In order of discovery:
PSR B1913+16: \citet{hulse75},
PSR B1820--11: \citet{Lyne89},
PSR B1534+12: \citet{Wolszczan91},
PSR B2127+11C: \citet{Prince91},
PSR J1518+4904: \citet{Nice96},
PSR J1811--1736: \citet{Lyne00},
PSR J0737--3039A/B: \citet{Burgay03,Lyne04},
PSR J1829+2456: \citet{Champion04},
PSR J1756--2251: \citet{Faulkner05}, and
PSR J1906+0746: \cite{Lorimer06}.
\end{center}
\end{table}

At present (2014) 7 DNSs, have been confirmed and 3 more are suspected (Table~\ref{tab_dns}).
Of these,  half have merger timescales shorter than a Hubble time. Half also show  eccentric
orbits, even at relatively short periods.
Bayesian statistical analyses based on these observations indicate that an optimistic Galactic DNS
merger rate may be up to $1.8~10^{-4}$
yr$^{-1}$, implying that their number should be approximately a few
million \citep{Kalogera01,Kalogera04} if we assume the age of the
Galaxy is $\sim14$ Gyr.  This number is roughly 1-2 orders of
magnitude higher than estimated from binary star population synthesis
\citep{Nelemans01b,Oslowski11,Dominik12}, although the uncertainty
in the Bayesian analysis can exceed one order of magnitude.

Theoretical investigations of the GW signal from Galactic DNSs have been carried out by ({\it e.g.})
\citet{Allen99,Belczynski10,Rosado11,Allen12}. The stochastic GW
background produced by DNS mergers in low-redshift galaxies
(up to z$\sim$5) has been investigated by \citet{Regimbau06} (also
see \citet{Rosado11}), who found that the signal should be
detectable by the new generation of ground-based interferometers.
\citet{zhu13} studied the GW background from compact binary mergers, and
showed that, below 100 Hz, the background
depends primarily on the local merger rate and the average
chirp mass and is independent of the chirp mass distribution. In
addition, the effects of cosmic star formation rates and delay times
between the formation and merger of binaries are linear below 100 Hz
in their model. \citet{Belczynski10} studied the GW background and
foreground signal from a Galactic  population of double compact
objects using the binary-star population-synthesis (BSPS) method.
They concluded that only a few (2-4) NS-NS binaries in
the Galaxy would have been detectable by the cancelled space observatory
LISA. However, approximations for (i) the calculation of the GW signal from
individual DNS binaries, and (ii) the employment of crucial initial conditions
and the treatment of important physical processes in the BSPS method, may
result in quite large uncertainties.

In order to understand the signal detected by sufficiently sensitive GW detectors,
it is necessary to characterize the radiation from all GW-emitting populations, including DNS.
This paper presents a study of the GW signal from the Galactic  population of {\it steady-state}
DNSs\footnote{ A DNS merger would be clearly indicated by a strong GW
signal at rapidly rising frequencies $>0.1$\,Hz. The most optimistic Galactic DNS merger rates are
$\simle10^{-3} {\rm yr^{-1}}$ -- see \S~\ref{sec_results}.
as a function of several key  parameters,  } including the Galactic star-formation history,
the initial-mass function, metallicity, and the physics of common-envelope evolution.
It examines how this DNS population differs from the
Galactic double-white-dwarf (DWD) population, and establishes a basis for computing the
GW signal due to extragalactic DNS populations.
We describe the methods used to model the DNS population in the Galactic disc,
the emission and superposition of GW signals, and the reduction of the data in terms of a
conceptual GW experiment  (eLISA) in  \S\,\ref{sec_method}.
Major results are presented in \S\,\ref{sec_results}.
Implications, observations and previous work are discussed
in \S\,\ref{sec_discussion}. The main conclusions are reviewed in \S\,\ref{sec_conclusion}.


\section{Methods}
\label{sec_method}

\subsection{Binary-star population synthesis}
\label{sec_bps}

In this section, we review the important physical processes
in binary star evolution and the initial and boundary conditions
for population synthesis to obtain a sample of double compact objects.
Population synthesis, including individual stellar-evolution tracks
and initial conditions, was carried out using the method described by
\citet{Yu10,Yu11} and \citet{Hurley00,Hurley02} in an initial study of
the present Galactic double degenerates population. Note that we take
some standard parameters for the Galactic disc (or the Galaxy) to
represent a Milky-way type galaxy.

\subsubsection{Common-envelope evolution}
\label{sec_ce}

When one star in a binary fills its Roche lobe either
by evolutionary expansion or by orbital shrinkage, Roche lobe
overflow (RLOF) occurs. The Roche radius of the primary is given
by
\begin{equation}
\frac{R_{\rm L_{1}}}{a}=\frac{0.49q_{1}^{2/3}}{0.6q_{1}^{2/3}+\rm
ln(1+\it q_{\rm 1}^{\rm 1/3} )}
\label{eq_rocherad}
\end{equation}
where $a$ is more generally the semimajor axis of the orbit and
$q_{1}(=m_{1}/m_{2})$ is the mass ratio of primary and secondary
\citep{Eggleton83}.
\citet{Hurley00,Hurley02} have shown that there is a critical
mass ratio $q_{\rm c}$ of a binary star which can be used to distinguish the stable RLOF
and common envelope (CE) phases, where $q_{\rm c}$ is a function of primary mass $m_1$,
its core mass $m_{\rm 1c}$, and the mass-transfer efficiency of the donor.
We adopt
\begin{equation}
q_{\rm c}=\left(1.67-x+2\left(\frac{m_{\rm 1c}}{m_1}\right)^{5}\right)/2.13 ,
\end{equation}
where $x$ = 0.3 is the exponent of the mass-radius
relation at constant luminosity for giant stars \citep{Hurley00,Hurley02}.

The calculation of the orbital parameters ({\it e.g.} orbital separation) of
a binary after CE ejection in our model is based on one of two  assumptions:
either 1) angular momentum conservation ($\gamma-$mechanism) or  2) energy conservation
($\alpha-$mechanism).

In the first case, we consider the angular momentum lost by a binary system
undergoing non-conservative mass transfer  to be described by the decrease of
primary mass times a factor $\gamma$
\citep{Paczynski67a,Nelemans00}:
\begin{equation}
\frac{J_{\rm i}-J_{\rm f}}{J_{\rm i}}=\gamma \frac{m_{1}-m_{1\rm
c}}{m_{1}+m_{2}},
\label{eq_am1}
\end{equation}
where $J_{\rm i}$ is the orbital angular momentum of
the pre-mass transfer binary; $J_{\rm f}$
is the final orbital angular momentum after CE ejection;
$m_{1}$ and $m_{1\rm c}$ is the primary mass and its core
mass respectively; $m_{2}$ is the secondary mass.

Combining the above equation with the fraction of angular
momentum lost during the mass transfer, $J_{\rm i} - J_{\rm f}$,
and Kepler's law, we have the ratio of final to initial orbital
separation
\begin{equation}
\frac{a_{\rm f}}{a_{\rm i}}=\left(\frac{m_{1}}{m_{1\rm
c}}\right)^{2}\left(\frac{m_{1\rm c}+m_{2}}{M}\right)\left(1-\gamma \frac{m_{1}-m_{1\rm
c}}{M}\right)^{2},
\label{eq_gamma}
\end{equation}
where $a_{\rm i}$ and $a_{\rm f}$ are the orbital separations before and
after the CE phase; $M=m_{1}+m_{2}$ is the sum of the primary and secondary
mass before the CE phase. \citet{Nelemans05} investigated the mass-transfer
phase of the progenitors of white dwarfs in binaries employing the
$\gamma$-mechanism based on 10 observed systems and deduced a
value of $\gamma$ in the range of 1.4 -- 1.7. In order to investigate
the influence of angular momentum loss on the rates of DCOs, we here adopt
$\gamma=$1.3 and 1.5.

In the second case, CE ejection of a binary star requires that the
envelope binding energy, including gravitational-binding and recombination energies,
must represent a significant fraction of the orbital energy \citep{Webbink84}.
\begin{equation}
\frac{G(m_{\rm 1}-m_{\rm 1c})m_{\rm 1}}
{\lambda r_{\rm L_{1}}}=
\alpha
\left(\frac{Gm_{\rm 1c}m_{2}}{2a_{\rm f}}-
      \frac{Gm_{1}m_{2}}{2a_{\rm i}}\right),
\end{equation}
where $\lambda$ is a structure parameter depending on the
evolutionary state of the donor, $\alpha_{\rm CE}$ is the CE
ejection efficiency representing how much orbital energy was
used to eject the CE, $r_{\rm L_{1}}$ is the Roche lobe radius
of the donor at the onset of mass transfer, and $G$ is the gravitational
constant \citep{Webbink84}. Rearranging in the form of Eq.\ref{eq_gamma},
we have
\begin{equation}
\frac{a_{\rm f}}{a_{\rm i}}=
\frac{m_{\rm 1c}}{m_{1}}
\left(1+\frac{2(m_{\rm 1}-m_{\rm 1c})a_{\rm i}}{\alpha\lambda m_{2}r_{\rm
L_{1}}}\right)^{-1}.
\label{eq_alpha}
\end{equation}
In this paper, we adopt $\lambda$ = 1.0, and $\alpha$ = 0.5 and 1.0.

A major difference between the  two CE ejection formulations is that energy
conservation implies a significant spiral-in stage in order
to eject the envelope (if we only consider the orbital energy as the
main engine) whilst the $\gamma-$ mechanism does not, implying that
the orbital separation can be larger after CE ejection in the latter
case. In particular, under the assumption of no external moment imposed
on a conservative mass-transfer binary, the angular momentum of the binary
must be conservative, and the final orbital separation can be written as
\begin{equation}
\frac{a_{\rm f}}{a_{\rm i}}=\left[\frac{(m_{1}-\Delta m)(m_{2}+\Delta m)}{m_{1}m_{2}}\right]^{2},
\end{equation}
where $\Delta m$ is the fraction of mass transferred from the primary to
the secondary. This means that in conservative evolution, if
$m_1 > m_2$  prior to mass transfer, the orbital separation incfreases
after mass transfer.

Although both CE ejection formulations can reproduce observations
\citep{Nelemans05,Webbink08} via a variation of free parameters,
considering both conservation laws may be a better approach to constrain
the CE evolution and final orbit of a binary.
Note that in this paper we neglect viscosity,  friction between the CE
and the stellar cores, and the potential nuclear and chemical energy in the
system.

\subsubsection{Gravitational radiation, magnetic braking, and tidal interaction}
\label{sec_orbit}

Other mechanisms which reduce the orbital separation of a binary system include
gravitational radiation and magnetic braking. A close compact binary system
driven by gravitational radiation may eventually undergo a mass transfer phase,
ultimately leading to coalescence. Using  the average energy ($E$) and angular momentum
($J_{\rm orb}$) loss during one orbital period,
we deduce the decay of orbital separation and eccentricity with respect to time to be
\begin{equation}
\frac{{\rm d} a}{{\rm d} t}=-\frac{64}{5}\frac{G^{3}m_{1}m_{2}(m_{1}+m_{2})}{c^{5}a^{3}}
\frac{1+\frac{73}{24}e^{2}+\frac{37}{96}e^{4}}{(1-e^{2})^{7/2}},
\label{eq_dota}
\end{equation}
\begin{equation}
\frac{{\rm d} e}{{\rm d} t}=-\frac{G^{3}m_{1}m_{2}(m_{1}+m_{2})}{c^{5}a^{4}}
\frac{\frac{304}{15}e+\frac{121}{15}e^{3}}{(1-e^{2})^{5/2}}.
\label{eq_dote}
\end{equation}
This calculation is consistent with our calculation of the GW signal from
DNS described in \S\,\ref{sec_gwnsns}.

Gravitational radiation could explain the formation of cataclysmic
variables (CVs) with orbital periods less than 3h, while magnetic
braking of the tidally coupled primary by its own magnetic wind
would account for orbital angular-momentum loss from CVs with
periods up to 10\,h \citep{Faulkner71,Zangrilli97}. We use the
formula for the rate of angular-momentum loss due to magnetic braking
derived by \citet{Rappaport83} and \citet{Skumanich72}:
\begin{equation}
\dot{J}_{\rm mb}=-5.83\times10^{-16}
\frac{m_{\rm env}}{m}
\left( \frac{r\omega_{\rm spin}}{\rm R_{\odot} yr^{-1}} \right)^{3}
~{\rm M_{\odot}R_{\odot}^{2}yr^{-2}},
\end{equation}
where $r$, $m_{\rm env}$ and $m$ are the radius, envelope mass and mass of
a star with a convective envelope, and $\omega_{\rm spin}$ is
the spin angular velocity of the star.

Tidal interaction caused by the gravity differential plays
an important role in the synchronization of stellar rotation and orbital
motion, and the circularization of the orbit. Relatively complete descriptions
of the tidal evolution have been given by \citet{Hurley02} and
\citet{Belczynski08}. In this paper, we adopt the same formulae and procedures
to deal with the tidal evolution as by \citet{Hut81,Zahn77,Campbell84,Rasio96} and \citet{Hurley02}.

\subsubsection{Formation of double neutron stars}

Our simulations assume three routes for the formation of neutron stars:
i) if a star has core mass of $m_{\rm c}\lesssim2.25 \Msolar$ at shell helium ignition,
it  evolve through double-shell thermal-pulses up the asymptotic giant branch. The star may
become a neutron star if its core mass grows and eventually exceeds
$2.25 \Msolar$; ii) if the core mass of the star on the thermal-pulsing
asymptotic giant branch does not exceed $2.25 \Msolar$ but it is heavy enough
($m_{\rm c}\gtrsim1.6 \Msolar$) to become an electron-degenerate oxygen-neon
white dwarf which may become a neutron star via accretion-induced collapse;
iii) if a star has a core mass of $m_{\rm c}\gtrsim 2.25 \Msolar$ at the start of the
early asymptotic giant (or red giant) branch, it will  become a neutron star without ascending
the thermal-pulsing asymptotic giant branch. If the core mass of a star at the time of supernova explosion
is sufficiently high ($\gtrsim7 \Msolar$), it will most likely become a
black hole unless significant mass loss takes place. These criteria are consistent
with  \citet{Hurley00}.

The gravitational mass of neutron stars is calculated by
\begin{equation}
m_{\rm ns}=1.17+0.09 m_{\rm \alpha},
\label{eq_massns}
\end{equation}
where $m_{\rm \alpha}$ represents either the mass of the carbon-oxygen
core at the time of supernova explosion or the mass of the oxygen-neon
core, estimated by $m_{\rm \alpha}=\max\{m_{\rm Ch},0.773m_{\rm c}-0.35\}$
with $m_{\rm Ch}$ being the Chandrasekhar mass.
Since $m_{\rm \alpha}$ is in the range of $\approx1.4-7 \Msolar$,
the masses of neutron stars are in the range of 1.3-1.8 \Msolar.
This is consistent with observational and theoretical constraints.
\citet{Lattimer07} show that about 83\% of observed neutron stars
have mass in the range $1-2 \Msolar$, while 100\% of observed
neutron stars have mass in the range $0.8-2.5 \Msolar$.
 For this paper, we assume the radius of a neutron star to be 10 km (\citet{Lattimer07}
give empirical values in the range 9 -- 15 km).

For the formation of DNS, we  assume seven evolution channels: \\
I:~ CE ejection $+$ CE ejection; \\
II:\, Stable RLOF $+$ CE ejection; \\
III: CE ejection $+$ stable RLOF; \\
IV:\, Stable RLOF $+$ stable RLOF; \\
V:~ Exposed core $+$ CE ejection;\\
VI:\, Solo CE ejection;\\
VII:\, Solo RLOF.

A binary with mass ratio $q=m_1/m_2$ less than some
critical value $q_{\rm c}$ will experience dynamically stable mass
transfer if the primary fills its Roche lobe while the star is in
the Hertzsprung gap or on the red giant branch. The
primary will become a compact object and the orbital separation will
change as
\begin{equation}
-{\rm d} \ln a =2 {\rm d} \ln m_{\rm 2}+ 2\alpha_{\rm RLOF} {\rm d}
\ln
 m_{\rm 1} + {\rm d} \ln (m_{\rm 1}+m_{\rm 2})
\end{equation}
where $\alpha_{\rm RLOF}$ is the mass-transfer efficiency for stable
Roche-lobe overflow (RLOF) \citep{Han95}. Here, we take $\alpha_{\rm
RLOF}=0.5$ \citep{Paczynski67a,Refsdal74}. Subsequently, if the {\it
secondary} fills its Roche lobe while it is in the Hertzsprung gap or
on the red giant branch, then RLOF will occur.

If the adiabatic response of the radius
of the mass donor is less than the change of its Roche lobe radius
with respect to a change of mass, {\it i.e.}
$\left(\frac{\partial \ln R_{\rm donor}}{\partial \ln M_{\rm donor}}\right)_{\rm ad}<
\left(\frac{\partial \ln R_{\rm RLOF}}{\partial \ln M_{\rm donor}}\right)_{\rm RLOF}$,
mass transfer will be unstable and a common envelope (CE) will form.
Interaction (friction) between the compact cores and the CE will
convert orbital energy into kinetic energy, heating and expanding
the CE. If the energy conversion mechanism is sufficiently
efficient, the CE will be expelled and a compact binary with a short
orbital period will result (see \S\,\ref{sec_ce}).

The above formation channels for DNSs represent the combination of RLOF
and CE processes. Note that channel V occurs when the envelope of a
massive primary is removed by a stellar wind rather than a first CE ejection.
CE ejection following evolution of the secondary may also give rise to a compact
binary.

\subsubsection{Neutron star kicks}
\label{sec_kicks}

Observations of proper motions indicate that pulsars have an extraordinary natal
velocity higher than their nominal progenitors \citep{Minkowski70,Lyne82,Lyne94,Hansen97,Fryer97}.
This may result from binary evolution \citep{Iben96} and asymmetric collapse and
explosion of supernovae \citep{Lai95,Lai01,Nordhaus12}. In this paper, we assume that
both cases can contribute to the acquisition of kick velocities by neutron stars.

Although \citet{Fryer97} and \citet{Arzoumanian02} suggested a possible bimodal distribution
for the kick velocities, we here simply assume that the kick velocities have a Maxwellian distribution
following the best estimate of \citet{Hansen97}, taking selection effects into account, with
\begin{equation}
\frac{{\rm d}N}{N{\rm d}v_{\rm k}}=(2/\pi)^{1/2}\frac{v_{\rm k}^{2}}{\sigma_{\rm k}^{3}}e^{-v_{\rm k}^{2}/2\sigma_{\rm k}^{2}},
\end{equation}
where $v_{\rm k}$ is the kick velocity and $\sigma_{\rm k}$ is its dispersion, ${\rm d}N/N$ is the normalized number
in a kick velocity bin ${\rm d}v_{\rm k}$.
We take $\sigma_{\rm k}=190$ km s$^{-1}$, so that the probable kick velocity $v_{\rm kp}\approx268$
km s$^{-1}$. As with other parameters in our population synthesis, a Monte Carlo
procedure is used to generate the individual kick velocities for the neutron stars.
Other parameters associated with the kick velocity (e.g. the direction)
are assumed to follow a uniform distribution.
For comparison, \citet{Hansen97} gives a kick velocity distribution of pulsars
with a mean value of about $250 - 300$ km s$^{-1}$ and $\sigma_{\rm k}=190$ km s$^{-1}$.
Without considering any selection effects, \citet{Lyne94} found a mean pulsar kick velocity
of 450$\pm$90 km s$^{-1}$. \citet{Hobbs05} suggested that there is lack of evidence for the
bimodal distribution of kick velocities and a $\sigma_{\rm k}=265$ km s$^{-1}$.

A kick adds a component to the orbital velocity  of neutron stars imposes,
leading to a change of the binary orbit.
We correct the values of orbital parameters of neutron stars using
Kepler$'$s laws and the binary dynamics. We have
\begin{equation}
a(1-e^{2})=\mid \textbf{d}\times \textbf{v}_{\rm o} \mid^{2}/(GM),
\label{eq_orb1}
\end{equation}
where $G$ is gravitational constant, $M$ the total mass of binary, $a$ the semi-major axis
of the orbit, $e$ the eccentricity, $\textbf{d}$ the distance vector between the two stars, and
$\textbf{v}_{\rm o}$ the orbital velocity of neutron star.
The orbital velocity $\textbf{v}_{\rm o}$ can be expressed as
\begin{equation}
\frac{1}{2}\textbf{v}_{\rm o}^{2} = GM\left(\frac{1}{\mid \textbf{d} \mid}-\frac{1}{2a}\right).
\label{eq_orb2}
\end{equation}
The velocity of the mass centre of the binary $\textbf{v}_{\rm c}$ relative to the old mass centre
can be simplified to
\begin{equation}
(M'-\Delta m'_{1})\textbf{v}_{\rm c}=m_{\rm 1c}\textbf{v}_{\rm k}+\Delta m'_{1}\frac{m'_{2}}{M'}\textbf{v}_{\rm o},
\label{eq_orb3}
\end{equation}
where $M'$ is the total mass of the binary before supernova collapse, $\Delta m'_{1}$ the mass loss
of the (exploding) primary due to the supernova collapse and explosion, and $m'_{2}$ the mass of secondary.
From Eqs.\,\ref{eq_orb1}-\ref{eq_orb2} and Kepler's laws, the new orbital parameters are determined.
We use the new orbital parameters to calculate the GW emission from DNS. Note that our model for
post-kick neutron stars is consistent with the dynamic model in \citet{Brandt95} and \citet{Hurley02}.

\subsubsection{Initial conditions for population synthesis}
\label{sec_ip}

In order to obtain a sample of compact binaries in the
Galactic thin disc which is comparable with observations,
we have performed a Monte-Carlo simulation in which we need the
six physical inputs described below. In this study,
we only investigate the effect of  the first three.
We use the different cases in our simulations to obtain
information on the population of DNS and their GW radiation in
other Galactic components via mass-scaling. The Galactic structure
is described in \S\,\ref{sec_structure}, and the results are presented
in \S\,\ref{sec_rgwgalaxy}.

\noindent (i)  We adopt three star formation (SF) models in our binary population synthesis
to see the influence of SF history on the population of compact binaries.
These are:\\
{\it Instantaneous SF:}  a single star burst  at the formation
of the thin disc with a constant SF rate of 132.9$M_{\odot}$ yr$^{-1}$ from 0 to 391 Myr.
followed by no SF from 391 Myr to 10 Gyr; \\
{\it Constant SF:}  SF occurs at a constant rate of
5.2$M_{\odot}$ yr$^{-1}$ from 0 to 10 Gyr;\\
{\it Quasi-exponential SF:} the star formation rate $S$
is the combination of a major star-forming process (the first term
of the following function) and a minor star formation (the second
term of the function),
\begin{equation}
S(t_{\rm sf})=7.92 e^{-(t_{\rm sf})/\tau}+0.09(t_{\rm sf})~~{\rm
    M_{\odot}yr^{\rm -1}}
\end{equation}
where $t_{\rm sf}$ is the time of star formation, and $\tau=9$ Gyr
\citep{Yu10}, which produces $\approx$3.5 $M_{\odot}$yr$^{-1}$ at
the current epoch.

All three models produce a thin-disc star mass of $\sim5.2\times10^{10}$
$M_{\odot}$ at the thin-disc age $t=10$ Gyr.
Observations place the current thin-disk SF rate in the range  $\approx 3 - 5
\Msolar {\rm yr^{-1}}$ \citep{Smith78,Timmes97,Diehl06}
and imply that {\it Instantaneous SF} is highly implausible for the
SFH in the thin disc. It is retained here for comparison.

\noindent (ii) The initial mass function (IMF) can be constrained by the local
luminosity function, stellar density and potential. The IMF for the
Galactic components may be different as indicated by \citet{Robin03},
\citet{Kroupa93} and \citet{Kroupa01} constrained by the observations
of \citet{Wielen83}, \citet{Popper80} and the Hipparcos mission
\citep{Creze98,Jahreiss97}.

In this paper, we adopt the frequently used power-law IMF
\begin{equation}
\xi(m)=~Am^{\sigma},~~0.1\leqslant m \leqslant 100 {\rm M_{\odot}}
\end{equation}
where $m$ is the primary mass; $\xi(m)\textrm{d}m$ is
the number of stars in the mass interval $m$ to $m+\textrm{d}m$;
$A$ is the normalization coefficient determined by
$A\int_{0.1}^{100}\xi(m)\textrm{d}m=1$. Since the results of
\citet{Kroupa93} and \citet{Kroupa01} indicate that $\sigma$
is in the range of $-1.3$ to $-2.7$, we take $\sigma=-1.5$ and $-2.5$
for comparison.

\noindent (iii) We have adopted a metallicity $Z=0.02$ (Population I) and 0.001
(Population II).

\noindent (iv) We assume a constant mass-ratio distribution
\citep{Mazeh92,Goldberg94},
\begin{equation}
n(1/q) = 1,  0.001 < 1/q < 1.
\end{equation}
The inverse mass ratio has a minimum value of 0.001
since the maximum and minimum mass of MS stars is 100 and 0.1 $M_{\odot}$
in our simulations.

\noindent (v) We employ the distribution of initial orbital separations used
by \citet{Han98} and \citet{Han03}, where they assume that all stars are
members of binary systems and that the distribution of separations
is constant in $\log a$ ($a$ is the separation) for wide binaries
and falls off smoothly at close separations:
\begin{equation}
\frac{{\rm d}a}{{\rm d}n}=\left\{
\begin{array}{c}
\alpha_{\rm sep}(\frac{a}{a_{0}})^{k},a\leqslant
a_{0},\\
\alpha_{\rm sep},a_{0}<a<a_{1}.
\end{array}
\right.
\label{eq_a}
\end{equation}
where $\alpha_{\rm sep}\approx0.070$, $a_{0}=10R_{\odot}$,
$a_{1}=5.75\times 10^{6}R_{\odot}=0.13$ pc, $k\approx1.2$. This
distribution implies that the number of binary systems per
logarithmic interval is constant. In addition, approximately 50 per
cent of all systems are binary stars with orbital periods of less
than 100 yr. These binaries are excluded when, during evolution,
the condition  that the sum of their radii exceeds their initial orbital separation is satisfied.

\noindent (vi) The distribution of initial eccentricities of binaries follows
$P_{e}=2e$ \citep{Heggie75,Nelemans01b}.

\subsubsection{Galactic mass distribution - potential, and rotational velocity}
\label{sec_structure}

\begin{table}
\begin{center}
\caption{Density laws and associated parameters. $r$ is the
spherical radius from the center of the Galaxy and $r_{0}$ is bulge
scale length; $R$ and $z$ are the natural cylindrical coordinates of
the axisymmetric disc, $h_{R}$ is the scale length of the disc,
$h_{z}$ is the scale height of the thin disc, $h'_{z}$ is the scale
height of the thick disc; $a'$ is the radius of the halo and $a'_{0}$
is a constant; $\rho_{c}$ is the central mass density. \label{tab_densitylaws}}
\scriptsize
\begin{tabular}{lccc}
\hline
&density law & constants   & $\rho_{c}$  \\
&            &  (kpc) & (M$_{\odot}\rm pc^{-3}$) \\
\hline
Bulge &$e^{-(r/r_{0})^{2}}$ &$r_{0}=0.5$ & $\frac{M_{\rm b}}{4\pi r_{0}^{3}}=12.73$ \\
\\
Thin disc &$e^{-R/h_{R}}\textrm{sech}^{2}(-z/h_{z})$ &$h_{R}=2.5$   & $\frac{M_{\rm tn}}{4\pi h_{R}^{2}h_{z}}=1.881$ \\
          &                                          &$h_{z}=0.352$ & \\
\\
Thick disc &$e^{-R/h_{R}}e^{-z/h'_{z}}$ & $h_{R} = 2.5$  & $\frac{M_{\rm tk}}{4\pi h_{R}^{2}h'_{z}}=0.0286$\\
           &                            & $h'_{z}=1.158$ & \\
\\
Halo &$ [(1+(\frac{a'}{a'_{0}})^{2})]^{-1}$ &$a'_{0} = 2.7$ & 0.108\\
\hline
\end{tabular}
\end{center}
\end{table}

We consider the Galaxy to comprise three components, namely the bulge, disc,
and a dark matter halo. We assume that the position of
the Sun is given by its distance ftom the Galactic centre $R_{\rm sun} = 8.5$\,kpc and height above
the Galactic plane $z_{\rm sun}$ = $16.5$\,pc
\citep{Freudenreich98}.
Our approximation for the Galactic density distribution is summarized in
Table~\ref{tab_densitylaws}. The detailed expressions are described as follows.

(1) We adopt a normal density distribution for the spherical bulge
with a cut-off radius of 3.5\,kpc \citep{Nelemans04},
\begin{equation}
\rho_{\rm b}(r) = \frac{M_{\rm b}}{4\pi r_{0}^{3}} e^{-(r/r_{0})^{2}}
    \quad {\rm M_{\odot} pc^{-3}},
\end{equation}
where $r$ is the radius from the center of the Galaxy,
$r_{0}= 0.5$\,kpc is the bulge scale length, and
$M_{\rm b}=2.0\times10^{10}$ is the mass of bulge. 
\citet{Robin03} suggest that the structure of the inner bulge
($<1^{\circ}$ from the Galactic center) is not yet well constrained
observationally. Consequently we here focus on the outer bulge and
make no allowance for any {\it additional} contribution to the
compact-binary population from the central region.

We use the potential proposed by \citet{Miyamoto75} in cylindrical coordinates
to calculate the rotational velocity of stars in the bulge.

(2) We model the thin and thick disc components of the
Galaxy using a squared hyperbolic secant plus exponential distribution
expressed as:
\begin{equation}
\rho_{\rm d}(R,z) = \frac{M_{\rm d}}{4\pi h_{R}^{2}h} e^{-R/h_{R}}\rho(z)
 \quad {\rm M_{\odot} pc^{-3}},
\end{equation}
where $R$ and $z$ are the natural cylindrical coordinates of the
axisymmetric disc, and $h_{R}=2.5$\,kpc is the scale length of the
disc, $h=h_{z}$ for the thin disc, $h=h'_{z}$ for the thick disc,
and $M_{\rm d}=M_{\rm tn}=5.2\times10^{10}~M_{\odot}$ is the mass of the thin disc;
$M_{\rm d}=M_{\rm tk}=2.6\times10^9~M_{\odot}$ is the mass of the thick disc.
$\rho(z)$ is the distribution in $z$, with:
\begin{equation}
\rho(z) = \textrm{sech}^{2}(-z/h_{z})~{\rm (thin~disc)}
\end{equation}
and
\begin{equation}
\rho(z) = e^{-z/h'_{z}}~{\rm (thick~disc)},
\end{equation}
where $h_{z}=0.352$\,kpc is the scale height of the thin disc and
$h'_{z}=1.158$\,kpc is the scale height of the thick disc. We neglect
the age and mass dependence of the scale height.

The Miyamoto \& Nagai potential in cylindrical coordinates is also used
to calculate the rotational velocity of stars in the disk.

(3)For the halo, we employ a relatively simple density distribution
which is consistent with \citet{Caldwell81},  \citet{Paczynski90} and \citet{Robin03}:
\begin{equation}
\rho_{\rm h}(a') =  \rho_{\rm c_h}
\left(1+\left(\frac{a'}{a'_{0}}\right)^{2}\right)^{-1},
\end{equation}
where $a'$ is the radius of the halo, $\rho_{\rm c_h}~=~0.108~M_{\odot}\rm pc^{-3} $
and $a'_{0}~=~2.7~{\rm kpc}$.

For the dark matter halo, we adopted the potential of \citet{Caldwell81}.
The observational estimate by \citet{Brand93} is included for comparison.

\begin{figure}
\centering
\includegraphics[width=10cm,clip,angle=0]{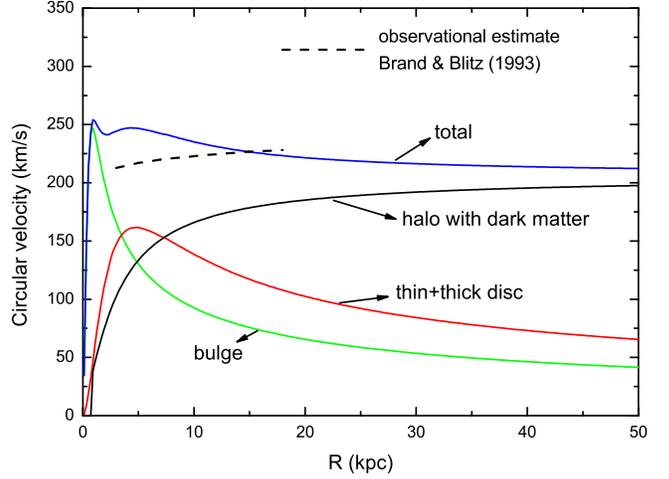}
\caption{ Rotational velocity as a function of galactocentric distance $R$
from the Galatic model, showing the contribution due to different
components,
{\it i.e.} the bulge, thin disc $+$ thick disc, and halo including
dark matter.
The dashed line indicates the observational estimate by \citet{Brand93}.
The spheroidal component due to the interstellar medium was not
considered separately.}
 \label{fig_rotation}
\end{figure}

Fig. \ref{fig_rotation} demonstrates the influence of the Galactic model on the rotation curve
of the Milky Way.

From the Galactic model, the total mass of the halo including
dark matter is $4.5\times 10^{11}{\rm M_{\odot}}$ inside a sphere of
radius 50\,kpc. We only focus on the baryonic mass in the halo which
is considered to be $5\times10^{10}{\rm M_{\odot}}$, constrained by the
density of double white dwarfs \citep{Yu10}.
With the SFR adopted here, the baryonic mass in the bulge and disc
is at least $2\times10^{10}{\rm M_{\odot}}$ and $5.5\times10^{10} {\rm
M_{\odot}}$ respectively, implying that our model assumes no dark matter
component localized to the bulge or the thin disc.

Combining the Galactic model and the mass of the Galactic components,
the stellar density in the solar neighbourhood is
$0.064 {\rm M_{\odot}pc^{-3}}$, of which
$6.27\times10^{-2}{\rm M_{\odot}pc^{-3}}$ is in the thin disc,
$9.4\times10^{-4}{\rm M_{\odot}pc^{-3}}$ is in the thick disc, and
$2.18\times10^{-5}{\rm M_{\odot}pc^{-3}}$ is in the halo.
This is consistent with the Hipparcos result, $0.076\pm0.015
{\rm M_{\odot}pc^{-3}}$ \citep{Creze98}.
The local dark matter density in our model is about
$0.01{\rm M_{\odot}pc^{-3}}$.

\subsection{Gravitational waves from double neutron stars}
\label{sec_gwnsns}

We calculate the GW strain amplitude from a single NS-NS
binary by linearizing the equations of general relativity
\citep{Peters63,Landau75}. We assume the following properties of a
binary. 1) The masses of the two components are $m_{1}$ and $m_{2}$
respectively. Therefore the total mass  $M=m_{1}+m_{2}$, and the
reduced mass $\mu=m_{1}m_{2}/M$. 2) The semi-major axis of the
orbit is $a$. 3) The eccentricity is $e$. 4) From Kepler, the orbital separation
$d=\frac{a(1-e^{2})}{1+e\cos(\varphi)}$ and the angular velocity
$\dot{\varphi}=\frac{(GM)^{1/2}[1+e\cos(\varphi)]^{2}}{a^{3/2}(1-e^{2})^{3/2}}$,
where $\varphi$ is the angle between the orbital separation and the
$x$-direction of an arbitrary Cartesian coordinate in the plane of the
binary orbit, and $G$ is the gravitational constant. We take the
$z-$direction perpendicular to the orbit plane and the origin is at
the center of mass. Clearly, $\int_{0}^{2\pi}{\rm
d}\varphi=\int_{0}^{P_{\rm orb}}\dot{\varphi}{\rm d}t$, where $P_{\rm
orb}$ is the orbital period.

The components of the strain amplitude can be expressed as
\citep{Landau75}
\begin{equation}
\left(
\begin{array}{ccccccccc}
h_{\rm xx} & h_{\rm xy} & h_{\rm xz}\\
h_{\rm yx} & h_{\rm yy} & h_{\rm yz}\\
h_{\rm zx} & h_{\rm zy} & h_{\rm zz}
\end{array}
\right)=-\frac{2G}{3c^{4}R_{\rm b}} \left(
\begin{array}{ccccccccc}
\ddot{A}_{\rm xx} & \ddot{A}_{\rm xy} & \ddot{A}_{\rm xz}\\
\ddot{A}_{\rm yx} & \ddot{A}_{\rm yy} & \ddot{A}_{\rm yz}\\
\ddot{A}_{\rm zx} & \ddot{A}_{\rm zy} & \ddot{A}_{\rm zz}
\end{array}
\right), \label{eq_saq}
\end{equation}
where $\ddot{A}_{\alpha\beta}$ represents the second order
differential of the mass-quadrupole tensor with respect to time,
suffix $\alpha\beta$ denotes the direction, $R_{\rm b}$ is the
distance form the observer, and $c$ is the speed of light. Since the mass
quadrupole tensor is
\begin{equation}
\begin{split}
&\left(
\begin{array}{ccccccccc}
A_{\rm xx} & A_{\rm xy} & A_{\rm xz}\\
A_{\rm yx} & A_{\rm yy} & A_{\rm yz}\\
A_{\rm zx} & A_{\rm zy} & A_{\rm zz}
\end{array}
\right)\\
&~~~~~~= \left(
\begin{array}{ccccccccc}
\mu d^{2}(3\cos^{2}\varphi-1) & 3\mu d^{2}\cos\varphi\sin\varphi & 0\\
3\mu d^{2}\sin\varphi\cos\varphi & \mu d^{2}(3\sin^{2}\varphi-1) & 0\\
0 & 0 & -\mu d^{2}
\end{array}
\right),
\end{split}
\label{eq_q}
\end{equation}
we have
\begin{equation}
\begin{split}
&\ddot{A}_{\rm xx} = -C_{2}6(\cos2\varphi+e\cos^{3}\varphi)+\ddot{A}_{\rm zz},\\
&\ddot{A}_{\rm yy} = C_{2}6(e^{2}+e\cos\varphi+\cos2\varphi+e\cos^{3}\varphi)+\ddot{A}_{\rm zz},\\
&\ddot{A}_{\rm zz} = -C_{2}2(e^{2}+e\cos\varphi),\\
&\ddot{A}_{\rm xy} = -C_{2}6(\sin2\varphi+2e\sin\varphi\cos^{2}\varphi+e\sin^{3}\varphi),\\
&\ddot{A}_{\rm yx} = \ddot{A}_{\rm xy}, \\
&C_{2}=\frac{GM\mu}{a(1-e^{2})}.
\end{split}
\label{eq_dqdt}
\end{equation}

From Eqs.\,\ref{eq_saq}, \ref{eq_q} and \ref{eq_dqdt}, it can be
shown that the GW from a binary with circular orbit is
monochromatic with frequency $2/P_{\rm orb}$. When the orbit is
eccentric, the wave becomes polychromatic and there is a frequency
broadening in the GW signal.

The average power of the GW radiated from two point masses over one
orbital period can be obtained by solving the third order
differential of the mass-quadrupole tensor with respect to time
\citep{Peters63,Landau75}. We quote the result
\begin{equation}
L_{\rm GW}=
\frac{32}{5}\left(\frac{G^{4}}{c^{5}}\frac{\mu^{2}M^{3}}{a^{5}}\right)\textit{z}(e)
\label{eq_lgw},
\end{equation}
\begin{equation}
z(e)=\frac{1+(73/24)e^{2}+(37/96)e^{4}}{(1-e^{2})^{7/2}}.
\end{equation}
After Fourier analysis of Kepler motion, we obtain the power in the
$n^{\rm th}$ harmonic \citep{Peters63}
\begin{equation}
L_{\rm GW}^{n}=
\frac{32}{5}\left(\frac{G^{4}}{c^{5}}\frac{\mu^{2}M^{3}}{a^{5}}\right)g(n,e)
\label{eq_lne},
\end{equation}
\begin{equation}
\begin{split}
g(n,e) &=
\frac{n^{4}}{32}\{[J_{n-2}(ne)-2eJ_{n-1}(ne)+\frac{2}{n}J_{n}(ne) \\
  & +2eJ_{n+1}(ne)-J_{n+2}(ne)]^{2} \\
  & +(1-e^{2})[J_{n-2}(ne)-2J_{n}(ne)+J_{n+2}(ne)]^{2} \\
  & +\frac{4}{3n^{2}}[J_{n}(ne)]^{2}\},
\end{split}
\label{eq_gne}
\end{equation}
where $J_n(ne)$ are Bessell functions of the first kind and
$n=1,2,3,...$. Since the sum of the power in each harmonic is equal
to the total power emitted from the binary, we have
\begin{equation}
\sum^{\infty}_{n=1}g(n,e)=z(e).
\end{equation}

After a mathematical transformation based on the equations above
\citep{Nelemans01b,Yu10}, we obtain the strain amplitude $h(n,e)$ in the vicinity of
the Earth at GW frequency $f_{n}$ in the $n^{\rm th}$ harmonic as
\begin{equation}
\begin{split}
h(n,e)& \equiv  h_{n} \\
& =4\sqrt{2}(2\pi)^{2/3}\frac{G^{5/3}}{c^{4}}M^{2/3}\mu P_{\rm
orb}^{-2/3}R_{\rm b}^{-1}\left(\frac{g(n,e)}{n^{2}}\right)^{1/2} \\
& = 1.14\times10^{-21} \\
& \times  \left(\frac{g(n,e)}{n^{2}}\right)^{1/2}
\left(\frac{\mathcal{M}}{M_{\odot}}\right)^{5/3} \left(\frac{P_{\rm
orb}}{\rm h}\right)^{-2/3} \left(\frac{R_{\rm b}}{\rm
kpc}\right)^{-1},
\end{split}
\label{eq_hne}
\end{equation}
\begin{equation}
f_{n} = n / P_{\rm orb}, \label{eq_fn}
\end{equation}
where $\mathcal{M}\equiv\mu^{3/5}M^{2/5}$ is the so-called {\it
chirp mass}. Eqs.\,\ref{eq_lne}, \ref{eq_gne}, \ref{eq_hne}, and
\ref{eq_fn} are the main equations used to calculate the power and strain
amplitude of the GW signal from one individual NS-NS pair in
frequency space. These equations also tell us that the power and
strain amplitude of the GW signal from a binary consisting of two
point-masses can be determined by four parameters - the chirp mass,
orbital period, eccentricity and distance. In this paper, we refer
to the first three as orbital parameters, and use the mass of each
component instead of the chirp mass.

The energy flux of GW waves can be expressed as
\begin{equation}
F=\frac{c^{3}\Omega^{2}}{16\pi G}h^2=\frac{c^{3}\pi f^{2}}{4 G}h^2,
\label{eq_flux}
\end{equation}
where $\Omega=2\pi f$ is the angular frequency.
We define the spectral function of the energy flux as
\begin{equation}
S=\frac{1}{\rho_{\rm c}c^{3}}\frac{{\rm d}F}{{\rm d}f}=\frac{2\pi^{2}}{3H_{0}^{2}}\frac{{\rm d}(f^{2}h^{2})}{{\rm d}f},
\label{eq_s}
\end{equation}
where $\rho_{\rm c}=3H_{0}^{2}/8\pi G$ is the critical mass
density of the present universe with $H_{0}\approx 73$ km s$^{-1}$ Mpc$^{-1}$
being the Hubble constant \citep{Freedman10}.
We show the GW energy flux spectrum for
selected cases to assist comparison with other work which uses this metric.

\subsection{The sensitivity of eLISA}
\label{sec_detector}

Evolved-LISA (eLISA) is  designed as an updated version of LISA,
a space-based GW detector, consisting of 1
mother and 2 daughter satellites flying in formation to form a
Michelson-type Laser interferometer with an arm length of
1$\times10^{6}$ km (see http://www.elisa-ngo.org/). Noise arises
mainly from the displacement noise (including the noise caused by
laser tracking system and other factors) and parasitic forces on the
proof mass of an accelerometer (acceleration noise)
\citep{Larson00}\footnote{The values of parameters to calculate the
total noise can be found on http://www.elisa-ngo.org/.}. We can
convert the noise signal to an equivalent GW signal in frequency
space by
\begin{equation}
h_{\rm f}=2\sqrt{\frac{S_{\rm n}}{R}},
\end{equation}
where $S_{\rm n}$ is the total strain noise spectral density,
$h_{\rm f}$ is the root spectral density and $R$ is the GW transfer
function given by \citet{Larson00}. In the simulations, we take the
displacement noise to be $1.1~10^{-11}$ mHz$^{-1/2}$ at 10 mHz, and
the acceleration noise to be $3~10^{-15}$ ms$^{-2}$Hz$^{-1/2}$ at 10
mHz. By comparison, the arm length of LISA would have been $5~10^{6}$ km. The
displacement noise and acceleration noise would have been  $4~10^{-11}$
mHz$^{-1/2}$ and $3~10^{-15}$ ms$^{-2}$Hz$^{-1/2}$.

For a continuous monochromatic source, such as a NS-NS binary with a
circular orbit, which is observed over a time $T$, the root spectral
density will appear in a Fourier spectrum as a single spectral line
in the form \citep{Larson00}
\begin{equation}
h_{\rm f}=h\sqrt{T}.
\end{equation}
So, for an observation time $T = 1$ yr, the root strain amplitude
spectral density $h_{\rm f}=5.62\times10^{3}h$ Hz$^{-1/2}$.

To demonstrate the detectability of the predicted GW signal due to the galactic
DNS population, we show the expected eLISA sensitivity for S/N=1 in
the figures in \S\,\ref{sec_results}.

\subsection{Data reduction}
\label{sec_reduction}

We reduce the simulated GW signal by using its mean integrated value $\langle h \rangle$.
To do this, we first choose a frequency interval
$\Delta f'$ which is greater than the interval used to compute the
simulations ({\it i.e.} $\Delta f=1 {\rm yr^{-1}}$). We then calculate the
mean value $\langle h \rangle$ of the strain amplitude and its standard
deviation $\sigma_{\langle h \rangle}$ in this large frequency interval using
\begin{equation}
\langle h \rangle=\frac{\sum^{j}_{i=1}h_{i}}{j},\\
\sigma^2_{\langle h \rangle}=\frac{\sum^{j}_{i=1}(h_{i}-\langle h \rangle)^{2}}{j},
\label{meansignal}
\end{equation}
where $j$ represents the number of small-frequency intervals in the large frequency interval.
In this paper, we take $\Delta \log f'=0.03$, so $j$ is also a function of frequency.
We plot $\langle h \rangle$ as a function of GW frequency in all figures unless specified
otherwise. In each panel, we also show the maximum standard deviation which represents the
maximum uncertainty in each large frequency interval in different cases.

\begin{table}
\caption{Main parameters and their values in our simulation.
See \S\,\ref{sec_bps} for an explanation of the parameters. }
\label{tab_pspace}
\begin{center}
\begin{tabular}{llllllllllccccccccccc}
\hline
\multicolumn{3}{c}{}&
\multicolumn{2}{c}{CEE($\alpha\lambda=$)}&
&
\multicolumn{2}{c}{CEA($\gamma=$)}\\
&&&       1.0  & 0.5 & & 1.3 &1.5 \\
 \hline
\multirow{7}{*}{Z=0.02}&SFH& IMF($\sigma=$)  &   &  & &  & \\
&Con& $-$1.5  & C1  & C3& & C25&C27 \\
&   & $-$2.5   & C2  & C4 & & C26&C28 \\
&Exp& $-$1.5  & C5  & C7& & C29&C31 \\
&   & $-$2.5   & C6  & C8 & & C30&C32 \\
&Inst& $-$1.5  & C9  & C11& & C33&C35 \\
&   & $-$2.5   & C10  & C12 & & C34&C36 \\
\hline
\multirow{7}{*}{Z=0.001}&SFH& IMF($\sigma=$)  &   &  & &  & \\
&Con& $-$1.5  & C13  & C15& & C37&C39 \\
&   & $-$2.5   & C14  & C16 & & C38&C40 \\
&Exp& $-$1.5  & C17  & C19& & C41&C43 \\
&   & $-$2.5   & C18  & C20 & & C42&C44 \\
&Inst& $-$1.5  & C21  & C23& & C45&C47 \\
&   & $-$2.5   & C22  & C24 & & C46&C48 \\
\hline
\end{tabular}
\end{center}
\end{table}

\subsection{Computation procedure}
\label{procedure}

In order to compute the superposition of the GW signal from the
entire DNS population in the Galactic disc and compare the signal
with the sensitivity of the proposed detectors, we need to know
their birth rates, merger rates, present number, space, mass,
eccentricity and orbital distributions. we have adopted the following
procedure to obtain the above physical properties.
For the Galactic thin disc with age  $t_{\rm disc}$, having a total mass
$M_{\rm tn}(t_{\rm disc})=\int_{0}^{t_{\rm disc}} {\rm S}(t_{\rm sf}) {\rm d}t_{\rm sf}$
\footnote{We here neglect the interstellar medium which makes up
about 20\% of the thin disc mass \citep{Robin03}.}:
\begin{enumerate}
\item For each star formation epoch in the disc, we calculate a sample
      distribution of $k$ coeval MS binaries having a total
      mass $m_{\rm p}$ and generated by the four Monte-Carlo simulation
      parameters $m$, $q$, $a$ and $e$.
\item We follow the evolution of each primordial binary in a time grid consisting
      of many time intervals to establish the properties of DNSs formed
      from the above MS binaries up to $t_{\rm disc}$.
       From the timescales from MS binary formation to DNS formation and from
      DNS formation to DNS merger, we can obtain the contribution function from
      each star formation epoch to the number of new-born and merged DNSs in a given
      time interval, as well as the total number of alive DNSs during that interval.
\item By summing the contributions from all star formation epochs in the thin disc,
      we can obtain all the physical information of the DNS population, including
       birth and merger rates, present number, and distributions of  orbital
       parameters.
\item Since most of the DNSs have eccentric orbits, we compute the Fourier
      transform of each orbit to obtain the value of GW strain emitted at each harmonic
      frequency  (as indicated by Eqs.\,\ref{eq_gne} and \ref{eq_hne}),
      and then sort the DNSs by the harmonics of orbital frequency (equivalent to the GW frequency).
\item We then calculate the total strain amplitude $h^2$ from the number and
      distance of DNSs in each frequency bin (for one year observation of e-LISA,
      see \S\,\ref{sec_detector}). This yields a raw data of strain amplitude against
      the GW frequency.
\item Finally, we use the method described in \S\,\ref{sec_reduction} to reduce the raw data, producing
      a reduced relation between the strain amplitude and GW frequency.
\end{enumerate}

In our population synthesis simulation, we started with 16 $10^{7}$ primordial MS binaries, distributed equally
between all 48 cases,
and assume that $t_{\rm disc}=10$ Gyr.
The parameters in the present study are SFH (Instantaneous, Continuous and Quasi-exponential),
IMF ($\sigma=-1.5$ and $-2.5$),
metallicity ($Z=0.02$ and 0.001),
and two different CE evolution processes ($\alpha$ and $\gamma$).
For each CE formalism we adopt two parameters ($\alpha\lambda=1.0$ and 0.5,
$\gamma=1.5$ and 1.3). The 48 individual cases are labelled
 C1--C48  and are summarized in Table\,\ref{tab_pspace}.


\begin{figure*}
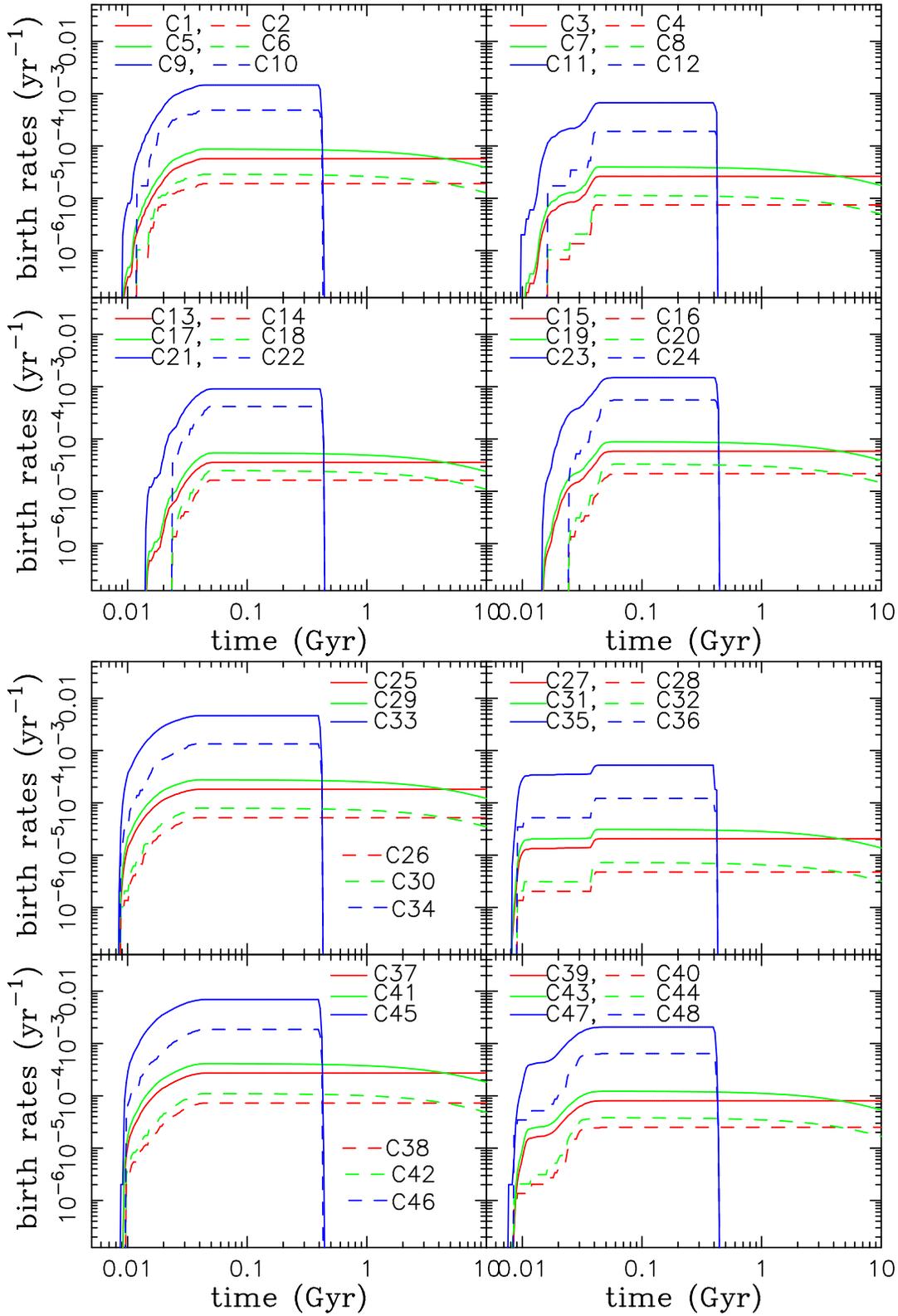

\centering
\subfigure
{
\begin{minipage}{5.9in}
\includegraphics[width=10.5cm,clip,angle=-90]{fig0.ps} \\
\includegraphics[width=10.5cm,clip,angle=-90]{fig1.ps}
\end{minipage}
}
\caption{Birth rates of DNSs in the Galactic disc in different cases.}
\label{fig_brates}
\end{figure*}

\begin{figure*}
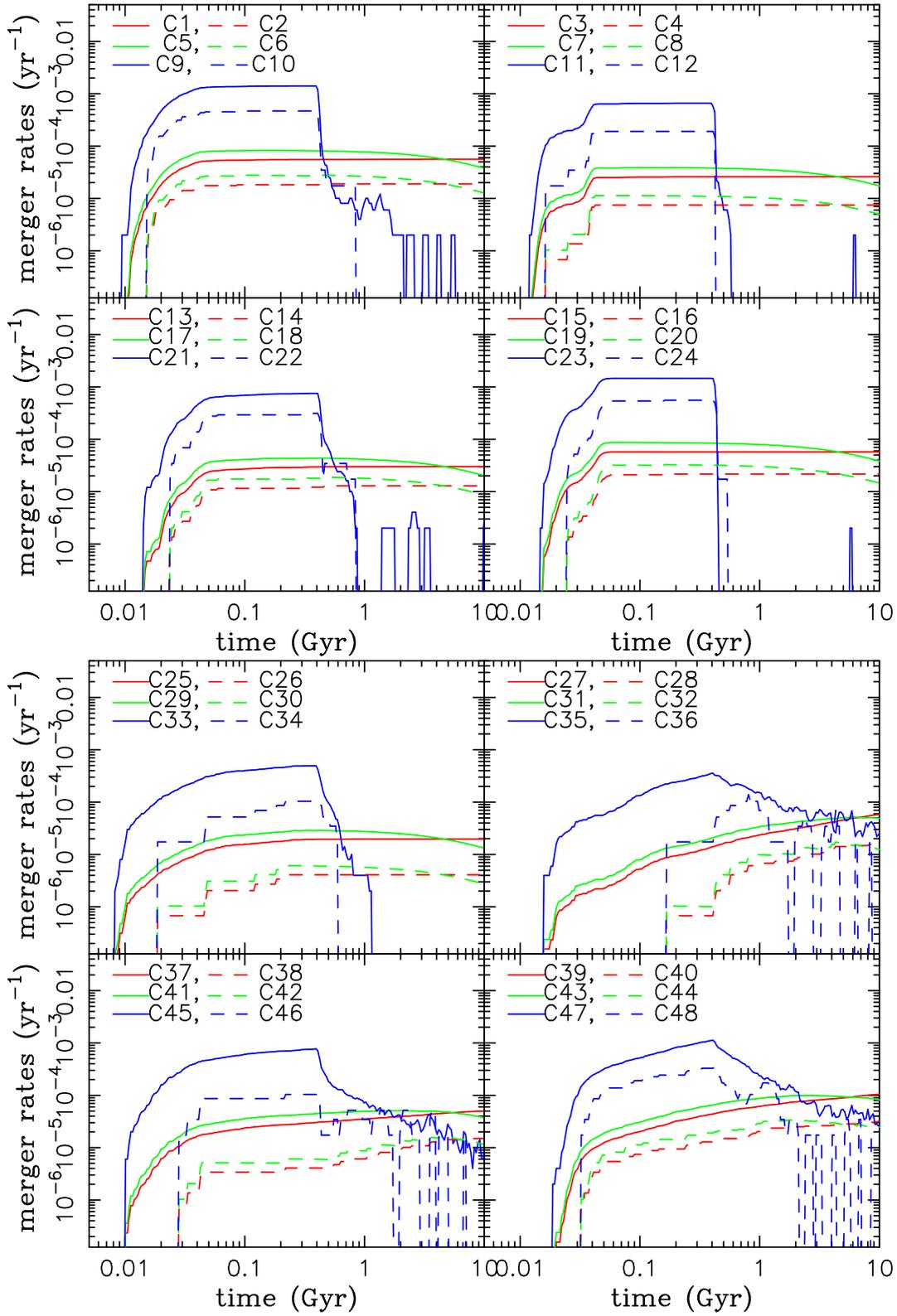

\centering
\subfigure
{
\begin{minipage}{5.9in}
\includegraphics[width=10.5cm,clip,angle=-90]{fig2.ps} \\
\includegraphics[width=10.5cm,clip,angle=-90]{fig3.ps}
\end{minipage}
}
\caption{Merger rates of DNSs in the Galactic disc in different cases.}
\label{fig_mrates}
\end{figure*}

\begin{table*}
\caption{Ranges in the present birth rate, merger rate and total number of DNS from binary-star population syntheses for a thin-disc age of 10\,Gyr.
Parentheses refer to simulation labels defined in Table~\ref{tab_pspace} and used in Figs. 1 -- 8. Merger rates and total numbers
may span a significant range, even when the present birth rate is zero, because of the delay time from new-born DNS to coalescence.}
\label{tab_rates}
\begin{center}
\begin{tabular}{lclll}
\hline
Star Formation & CE Ejection & Present Birth Rate & Present Merger Rate & Total Number \\
                       &                          &  $\times10^{-5}$yr$^{-1}$ & $\times10^{-5}$yr$^{-1}$ &     \\
\hline
Continuous & $\alpha$     & $0.49 - 5.8$ (C8 - C15) & $0.49 - 5.7$ (C8 - C15) & $61 - 5.5\times10^4$ (C20 - C13) \\
Continuous & $\gamma$  & $0.31 - 27$ (C32 - C37) & $0.27 - 11$ (C32 - C37) & $7.1\times10^3 - 1.9\times10^6$ (C32 - C37) \\
Instantaneous & $\alpha$ & $0$                                & $0$                                  & $0 - 5.2\times10^4$ (C10,12,24 - C21) \\
Instantaneous & $\gamma$ & $0$                             & $0 - 2.8$ (C34,36,48 - C45) & $ 6.4\times10^3 - 1.6\times10^6$ (C36 - C45) \\
\hline
\end{tabular}
\end{center}
\end{table*}

\section{Results}
\label{sec_results}

\subsection{The double neutron-star population}

\subsubsection{Birth rates, merger rates, and numbers}

The total number of DNS is a crucial input parameter for calculating the integrated
GW signal from the Galactic DNS population. The evolution of the birth and merger
rates of DNS in the thin disc are shown in Figs.\,\ref{fig_brates} and Figs.\,\ref{fig_mrates}.
There is a significant  difference in total DNS numbers between the instantaneous
SF model and the two other SF models. Since we assume that the response of descendent
stars to a normalized star-formation epoch with the same input parameters is an intrinsic
property,  there should be a quasi-linear relation between the SF rate and the DNS birthrate.
This can explain the different DNS birthrates in the three SF models.
For example, the maximum DNS birthrate (1.46 $10^{-3} {\rm yr^{-1}}$)
in case C9 (instantaneous SF: SFR$=133 \Msolar {\rm yr^{-1}}$)
is about 26 times that (5.72 $10^{-5} {\rm yr^{-1}}$) in the case C1 (constant SF: SFR$=5 \Msolar {\rm yr^{-1}}$).
The IMF affects the DNS birthrates in a similar way. Since the total number of initial MS
binaries in the sample is constant ($10^{7}$),
a small value  of the power-law index $\sigma$ (i.e. $-1.5$) increases the number of massive MS binaries and
hence gives higher DNS birthrates.

Although metallicity and CE ejection coefficients have no direct impact on the initial
number of MS binaries, they can significantly affect the orbital separation of a binary
following a CE phase. If, after first CE ejection, the orbital separation is sufficiently small,
the binary will not survive to become a DNS; this consequently reduces the DNS birthrate.
However, if a binary can survive both first and second CE ejection, there is an increased probability
of forming a very short-period compact binary. From Eqs.\,\ref{eq_gamma} and
\ref{eq_alpha}, either increasing $\gamma$ or decreasing $\alpha$ can, in principle,
lead to small orbital separation after CE ejection, and lower the DNS birthrate.
However, in the $\alpha-$formalism, this is not true for low metallicity ($Z=0.001$) ,
because the envelope of stars with low $Z$ have smaller radii and  usually have
a smaller energy absorption, resulting in smaller mass loss via weaker
stellar wind and in the formation of a massive core. From Eq.\,\ref{eq_alpha},
a larger core mass helps a  binary to survive, and thus increases the probability of
DNS formation. In summary, under the $\alpha-$formalism,  a lower metallicity leads to larger
core-mass and larger orbital separation, and increases the  DNS formation probability.
Decreasing the CE ejection coefficient $\alpha$ results in a smaller orbital separation and
lowers the DNS formation probability. In the $\gamma-$formalism, the situation is not so
obvious, but the orbital separation of a binary after a CE phase is also associated with the
core-mass of primary. The DNS birthrate depends on a competition between the metallicity
and the CE ejection coefficients.

Once a DNS has formed, its orbital separation and evolution is controlled
by  gravitational radiation and the magnetic field. The time scale for a MS binary to become
a DNS is generally  $\approx 10-40$ Myr. The  merger time  depends on the initial
 properties of the DNS ({\it i.e.}  mass, orbital period, and eccentricity).
Assuming the DNS orbital-period number distribution to be constant in logarithm,
(${\rm d}N/{\rm d}a\propto a^{-1}$), and the lifetime of an individual DNS to be $t\propto a^{4}$
({\it i.e.} ${\rm d}a/{\rm d}t\propto a^{-3}$),  then the lifetime number distribution
${\rm d}N/{\rm d}t\propto t^{-1}$.

\begin{figure*}
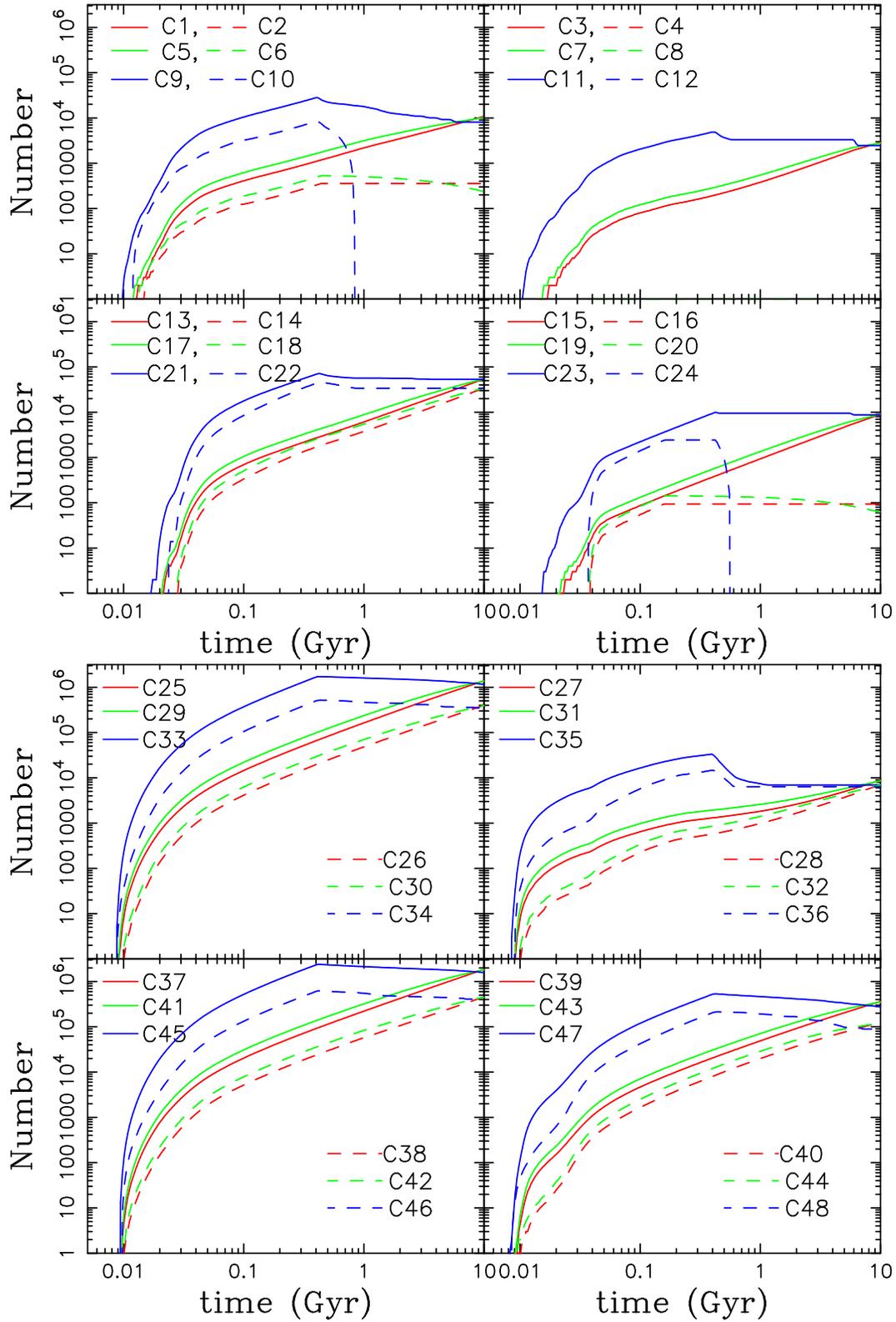

\centering
\subfigure
{
\begin{minipage}{5.9in}
\includegraphics[width=10.5cm,clip,angle=-90]{fig4.ps}\\
\includegraphics[width=10.5cm,clip,angle=-90]{fig5.ps}
\end{minipage}
}
\caption{The history of number of DNSs in the Galactic disc in different cases.
Note that the alive number of DNSs in cases C4, C8, and C12 is zero so we can
not see lines for them in this figure.  }
\label{fig_number}
\end{figure*}

However, we argue that the  DNS orbital-period
number distribution is {\it not simply} a constant in logarithm.
We compute the number evolution of the DNS population (see Fig.\,\ref{fig_number}) using two
methods to ensure consistency.
First,  we count the number from each star formation epoch. Second we calculate the
integral of the difference between birth rate and merger rate.
Our results show that the number of DNS in general depends on the total
evolutionary mass of the thin disc. Small differences in the predicted number at the
present epoch are found by using different metallicity and CE ejection coefficients.

Ranges in the current birth rates, merger rates and numbers of Galactic DNS assuming a thin-disc age of
10\,Gyr as given by our binary-star population-synthesis calculations are shown in Table \ref{tab_rates}.

We  draw attention to case C12 ($Z=0.02, \alpha\lambda=0.5$) in which, in our calculation
and as a consequence of  having orbital periods $\lesssim6000$ s,
DNSs can be born and merge within the same computational time interval.
This phenomenon leads to the total number of  DNS  always being zero in Fig.\,\ref{fig_number}
(there is no blue dashed line in  the panel showing C12).
The similar cases C10 and C24  can also produce DNSs with
short orbital periods ($\lesssim10^{6}$ s,  but longer than in C12). The DNSs in these two
cases  also merge quickly, resulting in zero DNSs at the present age of the thin disc (10 Gyr).
For comparison, as many as $\sim 10^{6}$  DNSs may be present at one time assuming a $\gamma-$mechanism,
that is  $1-10^{6}$ times higher than assuming an $\alpha-$mechanism. Even in the extreme case of instantaneous SF,
the present number can be $\sim 10^{4}-10^{6}$. This indicates that the $\gamma-$mechanism
can produce DNSs with orbital periods much longer than $\alpha-$mechanism.

\begin{figure*}
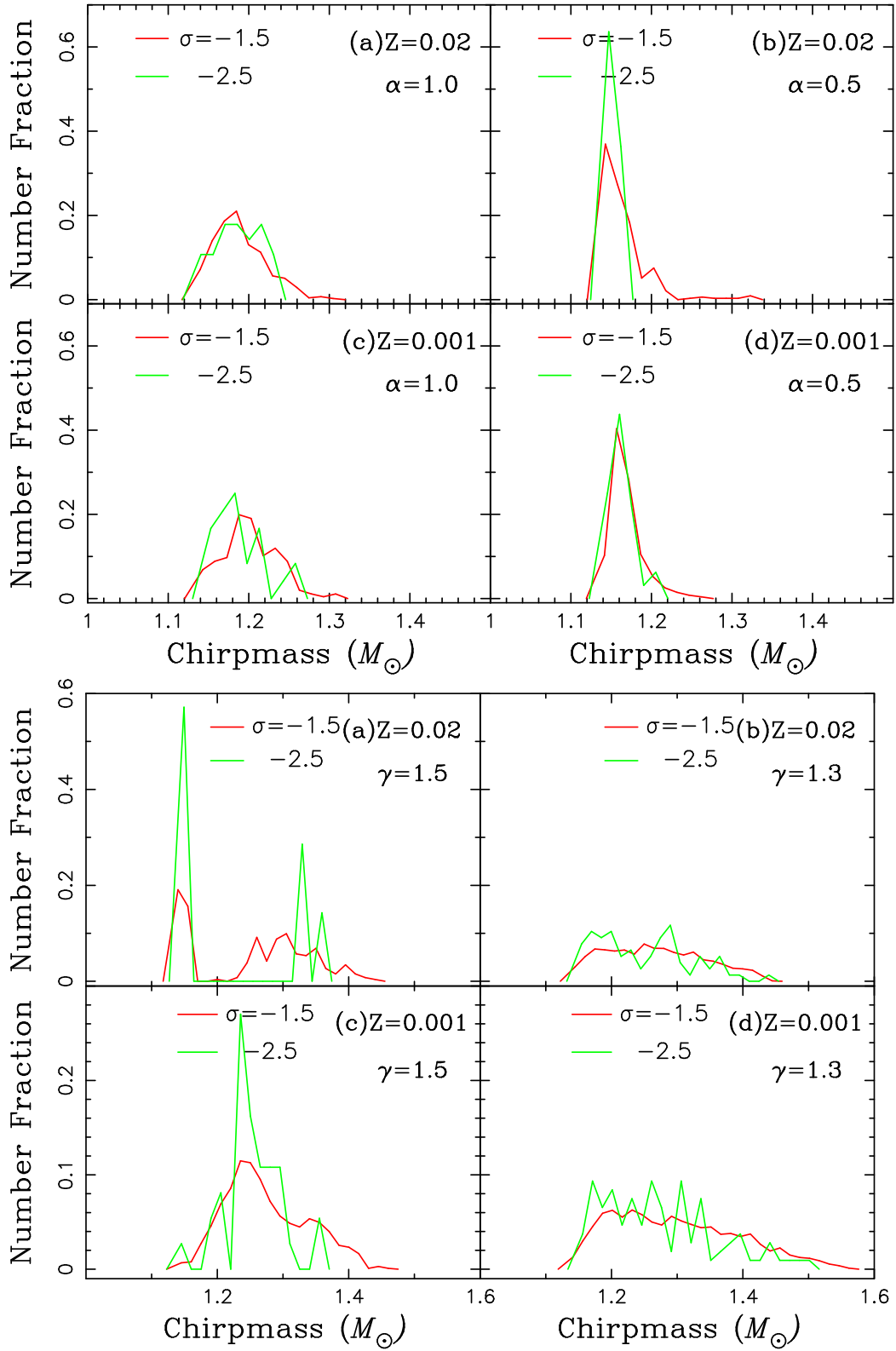

\centering
\subfigure
{
\begin{minipage}{5.9in}
\includegraphics[width=10.5cm,clip,angle=-90]{fig6.ps} \\
\includegraphics[width=10.5cm,clip,angle=-90]{fig8.ps}
\end{minipage}
}
\caption{The distribution of chirp mass of new-born DNSs
in the Galactic disc in different cases.
The number distributions (y-scale) are normalized to the unity. }
\label{fig_chm}
\end{figure*}

\begin{figure*}
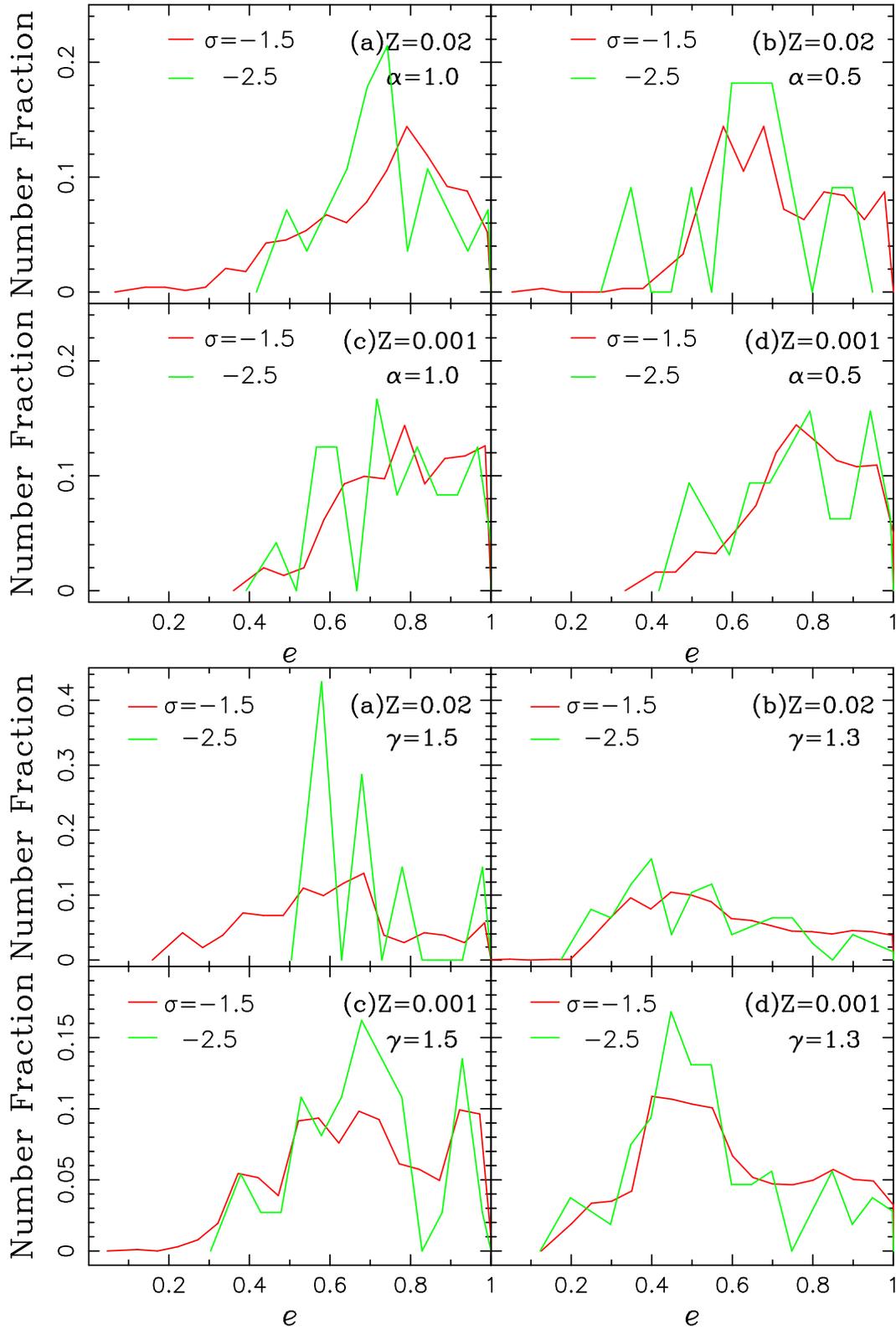

\centering
\subfigure
{
\begin{minipage}{5.9in}
\includegraphics[width=10.5cm,clip,angle=-90]{fig7.ps}\\
\includegraphics[width=10.5cm,clip,angle=-90]{fig9.ps}
\end{minipage}
}
\caption{The distribution of eccentricity of new-born DNSs
in the Galactic disc in different cases.
The number distributions (y-scale) are normalized to the unity. }
\label{fig_ecc}
\end{figure*}

\subsubsection{Chirpmass and eccentricity}

The distributions of chirp mass and eccentricity of the {\it new-born} DNS population are plotted in
Figs.\,\ref{fig_chm} and \ref{fig_ecc}. Since our calculation is based on a number of coeval MS stars and SF has
an overall effect on the calculation, the distributions of the physical variables of the DNS population
 ({\it i.e.}  mass, eccentricity, and orbital period) are here only influenced by the metallicity,
IMF and CE ejection coefficient.
Since the masses of neutron stars described in Eq.\,\ref{eq_massns} in our calculation are in the
range $1.3-1.8\,M_{\odot}$, we expect their chirp masses (see Eq.\,\ref{eq_hne}) to lie in the
range $1.1-1.6\,M_{\odot}$, as shown.
Our results show that, in general,  a lower $Z$ or higher IMF index increases the fraction of DNS with higher
chirp mass. Lower $Z$ leads to less mass loss and eventually leaves a more massive core, and
while a higher IMF index leads to more massive MS stars. In addition, we find
that lower $\alpha$ or higher $\gamma$ tend to give a more concentrated distribution at smaller
chirp mass. With no tidal interactions, the eccentricity distribution of the {\it new-born} DNS population
is centered around $0.6-0.8$ assuming an $\alpha-$mechanism and around $0.5-0.7$ assuming a
$\gamma-$mechanism. However, the eccentric orbits of the DNSs are likely to be circularized by
gravitational wave radiation and/or tidal interaction during the common-envelope phase and subsequent evolution.

\subsubsection{Orbital periods}

\begin{figure*}
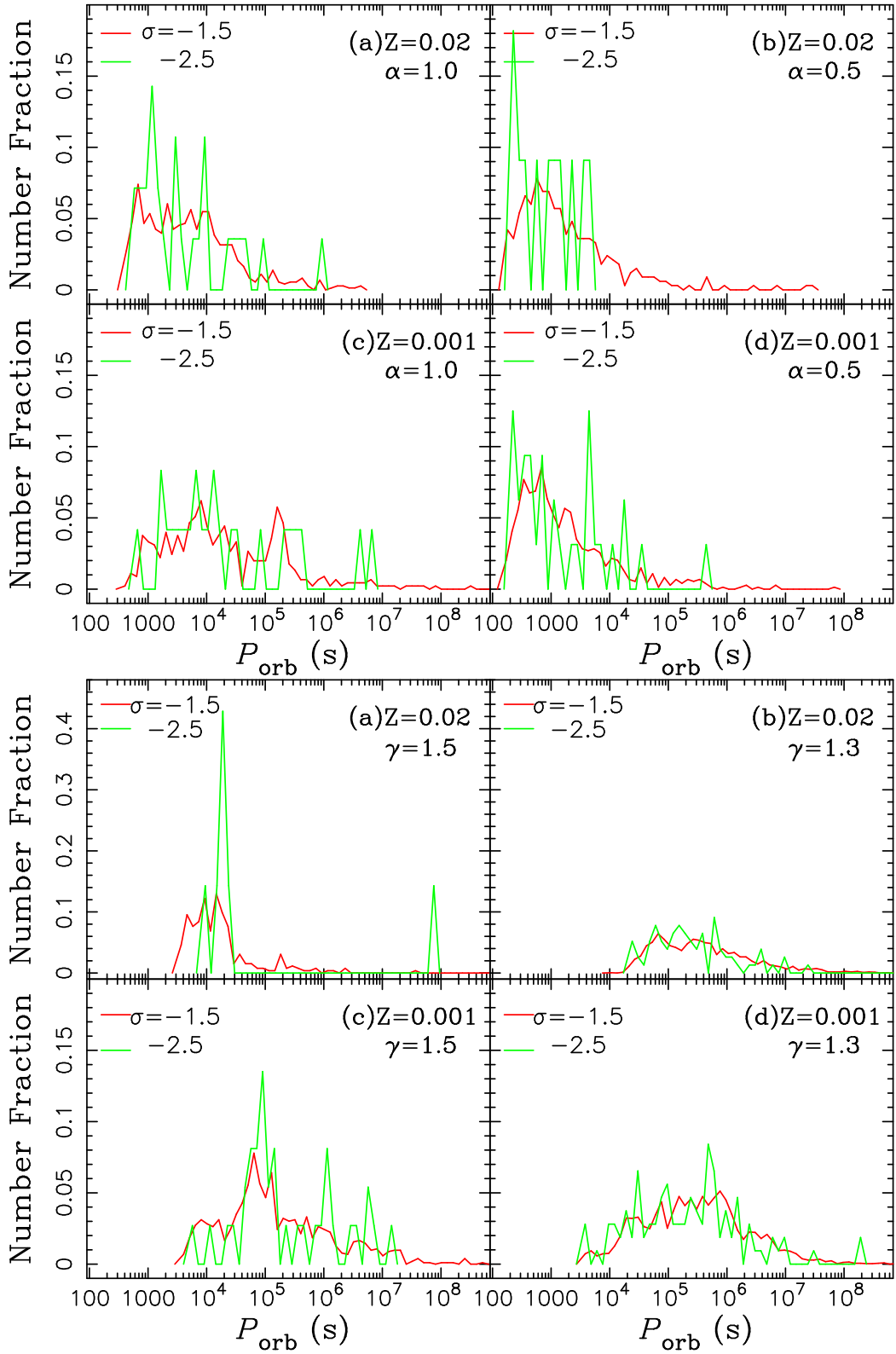

\centering
\subfigure
{
\begin{minipage}{5.9in}
\includegraphics[width=10.5cm,clip,angle=-90]{fig10.ps}\\
\includegraphics[width=10.5cm,clip,angle=-90]{fig11.ps}
\end{minipage}
}
\caption{The distribution of orbital periods of new-born DNSs
in the Galactic disc in different cases.
The number distributions (y-scale) are normalized to the unity. }
\label{fig_orb}
\end{figure*}

The difference in the orbital period distributions due to the two CE ejection models is
shown in Fig.\,\ref{fig_orb}. For the $\alpha-$mechanism, most  new-born
DNSs have short orbital periods with a peak in the range $200-10\,000$\,s. Increasing the IMF index
and/or $Z$ gives more DNSs with longer orbital periods. The shortest orbital
period in this model is about 200\,s ($\alpha=0.5$, $\sigma=-2.5$), while the longest
can be more than $10^{8}$\,s ($\alpha=1.0$, $\sigma=-1.5$, $Z=0.001$).
As argued already, the $\gamma-$mechanism yields DNSs with longer orbital periods.
The shortest orbital period in this model is about 3\,000\,s. In many cases
the new-born DNSs can have  orbital periods longer than $10^{8}$ s. These results are
consistent with the values of birth and merger rates in Figs.\,\ref{fig_brates} and \ref{fig_mrates}.

\subsubsection{Formation channels}

The fractions of DNSs  from different binary-star formation channnels are summarized for each model in Table~\ref{tab_channels}.
For $\alpha$-mechanism models it is found that most short-period DNSs ($P_{\rm orb}\lesssim10^{4}$ s)
come from the CE+CE channel, while longer period DNSs generally come  from channels with at least once stable RLOF
process.  The dramatic decrease in the number of DNSs with short orbital periods ($\lesssim10^{4}$ s) in the
$\gamma$-mechanism models is most likely due to the decrease of number of double
CE ejection events. We find that the number fraction of DNSs with short orbital periods may be up
to 10\% in the two $\gamma$-mechanism models where double-CE events contribute over 80\% of the DNSs.

\begin{table}
\caption{The fraction of DNSs from different formation channels, expressed as a percentage,
assuming  constant star-formation.  }
\label{tab_channels}
\begin{center}
\begin{tabular}{ccccccc}
\hline
\multicolumn{2}{l}{CE}                     & $Z$   & $\sigma$ & CE+CE & RLOF+CE & other  \\
\multicolumn{2}{l}{Ejection}             &          &                &               & CE+RLOF &      \\
 \multicolumn{2}{l}{Model}              &          &                &       \%      &     \% &   \%     \\
\hline
   &  $\alpha$ \\
C1   & 1.0 & 0.02   & $-1.5$ & 87.2  &  1.6 & 11.2   \\
C2   & 1.0 & 0.02   & $-2.5$ & 82.1  &  0.0 & 17.9   \\
C3   & 0.5 & 0.02   & $-1.5$ & 93.6  &  1.0 & 5.4   \\
C4   & 0.5 & 0.02   & $-2.5$ & 100.0  &  0.0 & 0.0   \\
C13  & 1.0 & 0.001  & $-1.5$ & 58.0  &  15.9 & 26.1     \\
C14  & 1.0 & 0.001  & $-2.5$ & 58.1  &  21.0 & 21.0  \\
C15  & 0.5 & 0.001  & $-1.5$ & 95.3  &  1.6 & 3.1   \\
C16  & 0.5 & 0.001  & $-2.5$ & 100.0  &  0.0 & 0.0   \\
\hline
 & $\gamma$ \\
C25   & 1.3 & 0.02   & $-1.5$ & 44.4  &  55.5 & 0.1   \\
C26   & 1.3 & 0.02   & $-2.5$ & 48.1  &  51.9  & 0.0   \\
C27   & 1.5 & 0.02   & $-1.5$ & 94.3  &  4.6 & 1.1   \\
C28   & 1.5 & 0.02   & $-2.5$ & 85.7  &  0.0 & 14.3   \\
C37   & 1.3 & 0.001  & $-1.5$ & 37.9  &  60.4 & 1.7     \\
C38   & 1.3 & 0.001  & $-2.5$ & 47.7  &  52.3 & 0.0  \\
C39   & 1.5 & 0.001  & $-1.5$ & 64.7  &  33.8 & 2.5   \\
C40   & 1.5 & 0.001  & $-2.5$ & 59.5  &  35.1 & 5.4   \\
\hline
\end{tabular}
\end{center}
\end{table}

\begin{figure*}
\centering
\subfigure
{
\begin{minipage}{5.9in}
\includegraphics[width=10.5cm,clip,angle=-90]{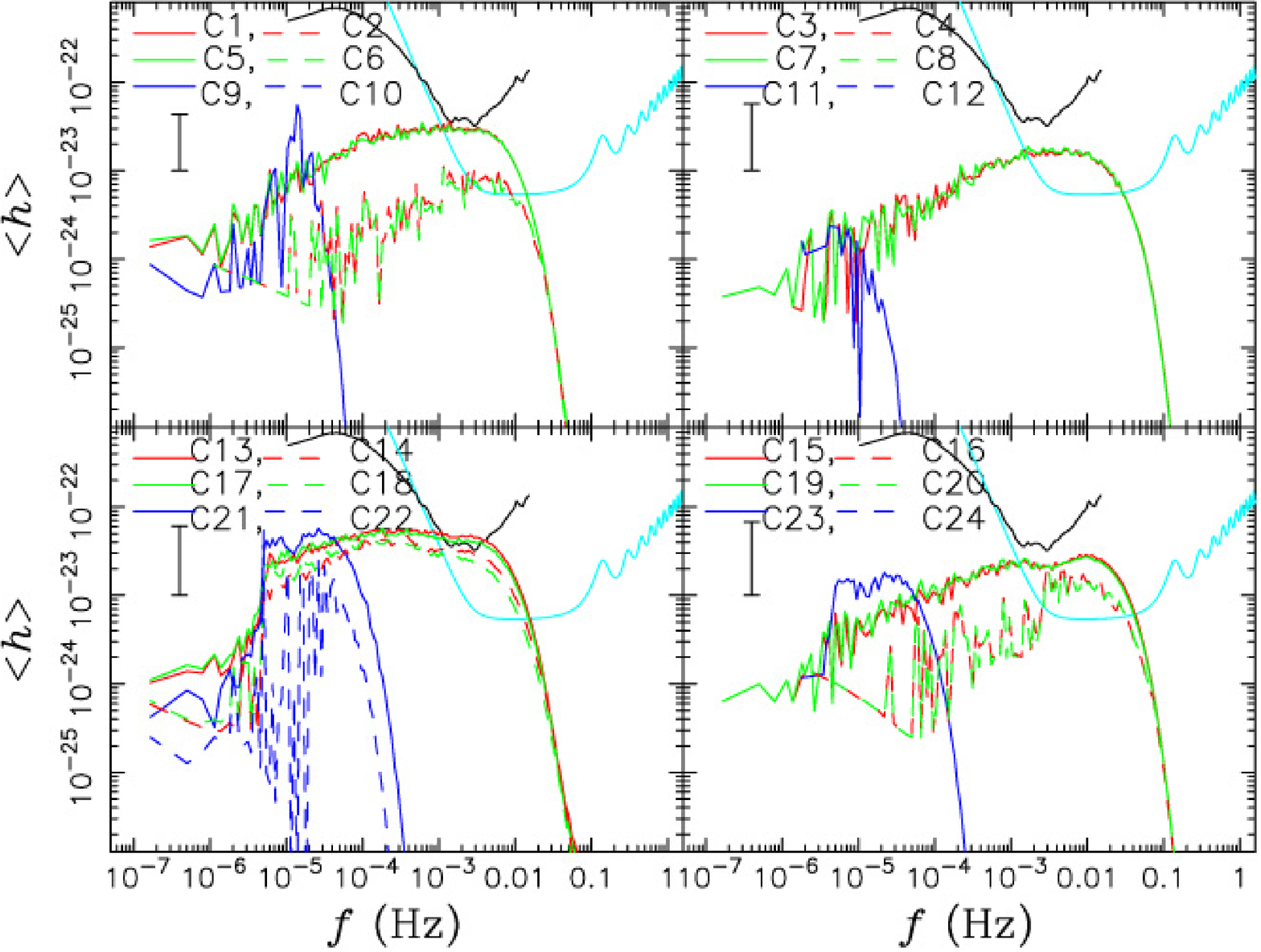}\\
\includegraphics[width=10.5cm,clip,angle=-90]{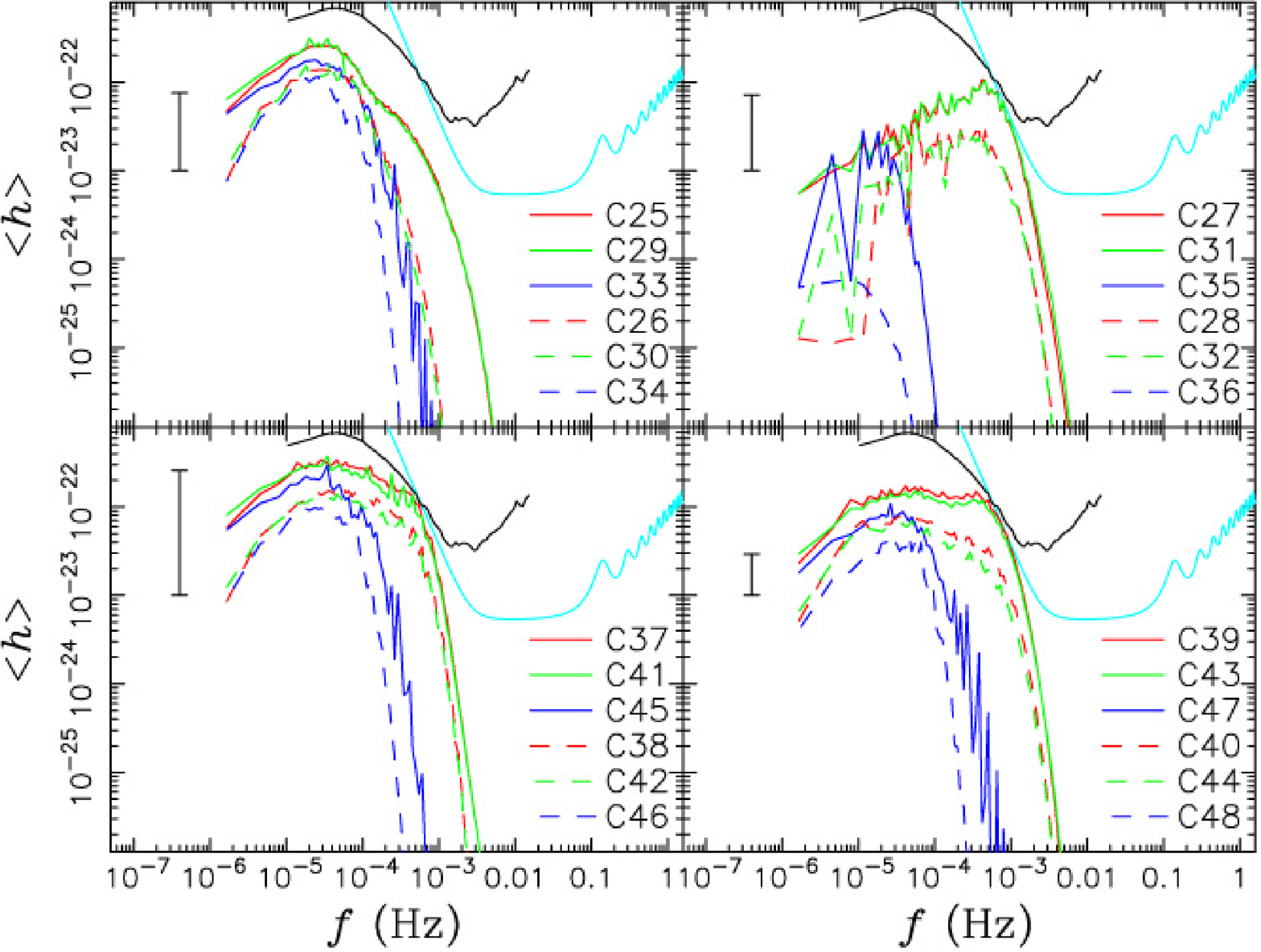}
\end{minipage}
}
\caption{The gravitational wave signal from a DNS population
in the Galactic disc for different cases. The black line shows
the gravitational wave noise due to double white dwarfs taken from \citet{Yu10}.
There is no gravitational wave signal from double neutron stars (DNS) in
cases C4, C8, C10, C12 and C24 since the current total number of DNS in these cases is zero.
The cyan lines show the sensitivity of eLISA at $S/N=1$.
The upper-left lines denotes the maximum deviation of the GW strain from the averaged value.
We take the bin size of $\Delta \log (f/{\rm Hz})$=0.03.
}
\label{fig_gwsignal}
\end{figure*}%

\subsubsection{The influence of kicks of neutron stars}
\label{sec_rkicks}

As initialized, the kick velocities of neutron stars
in our simulations obey a Maxwellian distribution
with a maximum likelihood (in 3D) being $\approx268$ km s$^{-1}$.
With a standard deviation of 190 km s$^{-1}$,
kick velocities  can lie in the range  10 -- 1200 km s$^{-1}$.
Since the new orbit of a binary is derived from the kick velocity (\S\,\ref{sec_kicks}), it has a
crucial influence on the distribution of eccentricity and orbital
period. Kick velocities mean that about 95\% of DNSs have
eccentric orbits,  with about 90\%  from supernova kicks,
and about 5\% from binary evolution. Our results indicate that
young DNSs may have highly eccentric orbits, although gravitational wave
radiation (and magnetic braking) tends to circularize the orbits during late evolution.
By comparison, all the binary white dwarfs in the simulations of \citet{Yu10}
have $e=0$ due to the tidal interaction and zero kick velocity.

Considering the extreme case of the neutron star velocity to be the sum
of the maximum kick velocity and the Galactic rotation velocity, i.e. 1200+250 km s$^{-1}$,
their maximum speed is about 1.5 kpc Myr$^{-1}$; the average will be closer to
268+250 km s$^{-1} = 0.54$ kpc Myr$^{-1}$.
The measured GW strain depends on the distance; the change in strain
over the course of one year due to neutron star kick velocities will be negligible.

\subsection{The GW strain-amplitude -- frequency relation}
\label{sec_hfrelation}

Fig.\,\ref{fig_gwsignal}.
shows the DNS GW signal for each model in Table\,\ref{tab_pspace} as a
strain- amplitude -- frequency relation, reduced using the method described
in \S\,\ref{sec_reduction}, so as to simplify the information contained.
The probability to detect a GW signal from a DNS population using e-LISA,
assuming an instantaneous SF model,  is very low or close to to zero.
If the SF rate is continuous or there is significant star formation in the recent past,
we expect a high probability to detect a GW signal from a DNS population, assuming
common-envelope ejection follows the $\alpha-$mechanism.
If CE ejection follows a $\gamma-$mechanism, DNS GW detectability with e-LISA  is also very low.
The $\gamma$-mechanism can produce more DNSs at frequencies, but only at frequencies  about 2 orders
of magnitude lower than the best working frequency of e-LISA ($10^{-2}$ Hz).

Most of the strain-amplitude relations show oscillations in $\langle h(f) \rangle$,
sometimes with amplitudes of more than 1 dex.
The DNS chirp masses in our calculation lie in the range  $1.1-1.6\, M_{\odot}$,
so can affect $\langle h \rangle$ by a factor of 1.87.
$\sim$95\% of DNSs in our model lie at distances between 3.92 and 13.92\,kpc,
affecting the signal by a factor of up to 3.55. These two parameters are unlikely
to explain the oscillation. The distribution of DNS orbital periods should
be the most important factor. This can be understood as follows.

We first assume that any probability distribution function can be
fitted by a normal  or multi-normal distribution, so
that the simplest distribution function is the normal distribution.
We apply this distribution to the orbital periods to understand the
$\langle h(f) \rangle$ relation. The frequency distribution can be expressed as
\begin{equation}
\frac{{\rm d}N}{{\rm d} f}=\frac{{\rm d}N}{{\rm d} P_{\rm orb}}\cdot
\frac{{\rm d}P_{\rm orb}}{{\rm d} f}=\sum^{n_{\rm
max}}_{n=1}nf^{-2}\cdot \frac{{\rm d}N}{{\rm d} P_{\rm orb}}.
\end{equation}
If the distribution of $P_{\rm orb}$ is normal, then
\begin{equation}
\frac{{\rm d}N}{{\rm d} f}=\sum^{n_{\rm max}}_{n=1}nf^{-2}\cdot
C_{\rm n}\exp\left(-\frac{(nf^{-1}-P_{\rm
orb,0})^{2}}{\sigma_P^{2}}\right). \label{eq_dndfnormal}
\end{equation}
To find the extrema $f_{\rm n}$ of the function to be summed in
Eq.\,\ref{eq_dndfnormal}, we set its first derivative with
respect to frequency $f$ equal to zero.
After rearranging,
\begin{equation}
f_{\rm n}^{2}+\frac{nP_{\rm orb0}}{\sigma^{2}}f_{\rm
n}-\frac{n^{2}}{\sigma_P^{2}}=0,
\end{equation}
where  $f_{\rm
n}=n\left[\frac{-\eta+\sqrt{\eta^{2}+4}}{2\sigma_P}\right]$ and
$\eta=P_{\rm orb,0}/\sigma_P$. If $\sigma_P \approx \eta$, the frequency
distribution may have $n_{\rm max}$ maximum values, and the interval
between two successive maxima is
$\left[\frac{-\eta+\sqrt{\eta^{2}+4}}{2\sigma_P}\right]$. If
$\sigma_P \gg \eta$, the interval becomes negligible.

The total strain amplitude $h_{\rm t}^{2}$ in one frequency bin has
been calculated as the sum of the strain amplitude $h^{2}$ of those
binaries in the frequency bin, expressed as
\begin{equation}
h_{\rm t}^{2}=\sum^{\frac{{\rm d}N}{{\rm d}f}\Delta f}_{N=1} h^{2}.
\end{equation}
If we assume the binaries in a frequency bin have similar chirp
mass, the above equation becomes
\begin{equation}
h_{\rm t}^{2}=C_{\rm h}\frac{{\rm d}N}{{\rm d}f}f^{4/3}
n^{-2}g(n,e),
\end{equation}
where $C_{\rm h}$ is a constant. Combining with
Eq.\ref{eq_dndfnormal}, we obtain an expression for the strain
amplitude generated from DNS binaries with normally distributed
 orbital periods:
\begin{equation}
h_{\rm t}^{2}=C_{\rm h}\sum^{n_{\rm
max}}_{n=1}n^{-1}f^{-2/3}\exp\left(-\frac{(nf^{-1}-P_{\rm
orb,0})^2}{\sigma_P^2}\right)g(n,e).
\end{equation}
To find the extrema, we take the
derivative  ${\rm d/d}f$ of the function in the sum and obtain
\begin{equation}
f_{\rm nh}^{2}+3\frac{nP_{\rm orb,0}}{\sigma^{2}}f_{\rm
nh}-3\frac{n^{2}}{\sigma_P^{2}}=0,
\end{equation}
which has the solutions
$f_{\rm nh}=n\left[\frac{-3\eta+\sqrt{9\eta^{2}+12}}{2\sigma_P}\right]$.
The properties of $f_{\rm nh}$ are similar to those of $f_{\rm n}$. In
addition,  when $\eta \gg 1$,  $f_{\rm hm}\approx f_{\rm m}$.

Consequently, if the orbital periods satisfy a perfect normal distribution,
an oscillation can appear in the $\langle h(f) \rangle$ relation under the condition
that neighboring frequencies coincide with extrema of $\langle h \rangle$.
However, since the orbital periods are most likely the sum of normal distributions
with different $P_{\rm orb,0}$ and $\sigma_{P}$, the frequencies where the strain
amplitudes have extrema relate to $\eta$ and $\sigma_{P}$.

This analysis demonstrates the following. If a population
of Galactic DNS has a relatively large distribution of orbital period (large $\sigma_P$),
it  will generate a smooth GW background. However, with small $\sigma_P$,
or a tight distribution of period,  we see a set of harmonics in the frequency range
$10^{-6}-10^{-2}$ Hz. All the DNSs in the sample are effectively``in tune".
With larger $\sigma_P$ this effect is smeared out. This means that under some conditions
({\it e.g.} $P_{\rm orb}/\sigma \gg 1$), the oscillation of space caused by the GW
radiation from a DNS population at low frequencies can be regularly enhanced.
We note that the oscillation in our simulations is hardly  detectable by eLISA.

For comparison, Fig.\,\ref{fig_gwsignal}  also shows the reduced GW signal from double white dwarfs (DWDs)
simulated by the method in \citet{Yu10}. The signals from DWDs occupy ($\lesssim$20\%) of the
observation frequency intervals for one year of observation, and they should have a different polarization
pattern from the signal from DNSs. This means that future observations should be
able to distinguish the DNS and DWD signals.

\begin{figure*}
\centering
\subfigure
{
\begin{minipage}{5.9in}
\includegraphics[width=10.5cm,clip,angle=-90]{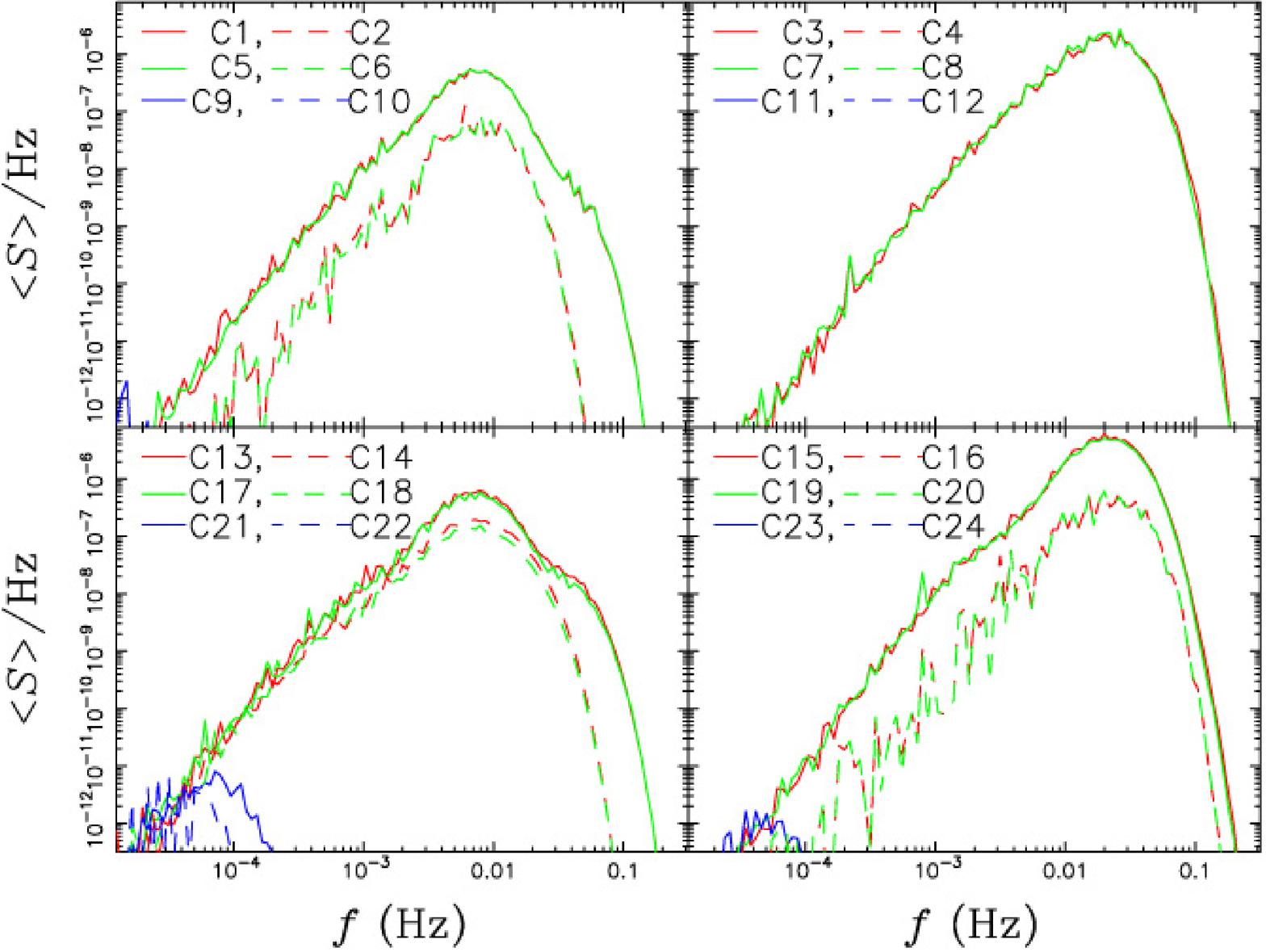}\\
\includegraphics[width=10.5cm,clip,angle=-90]{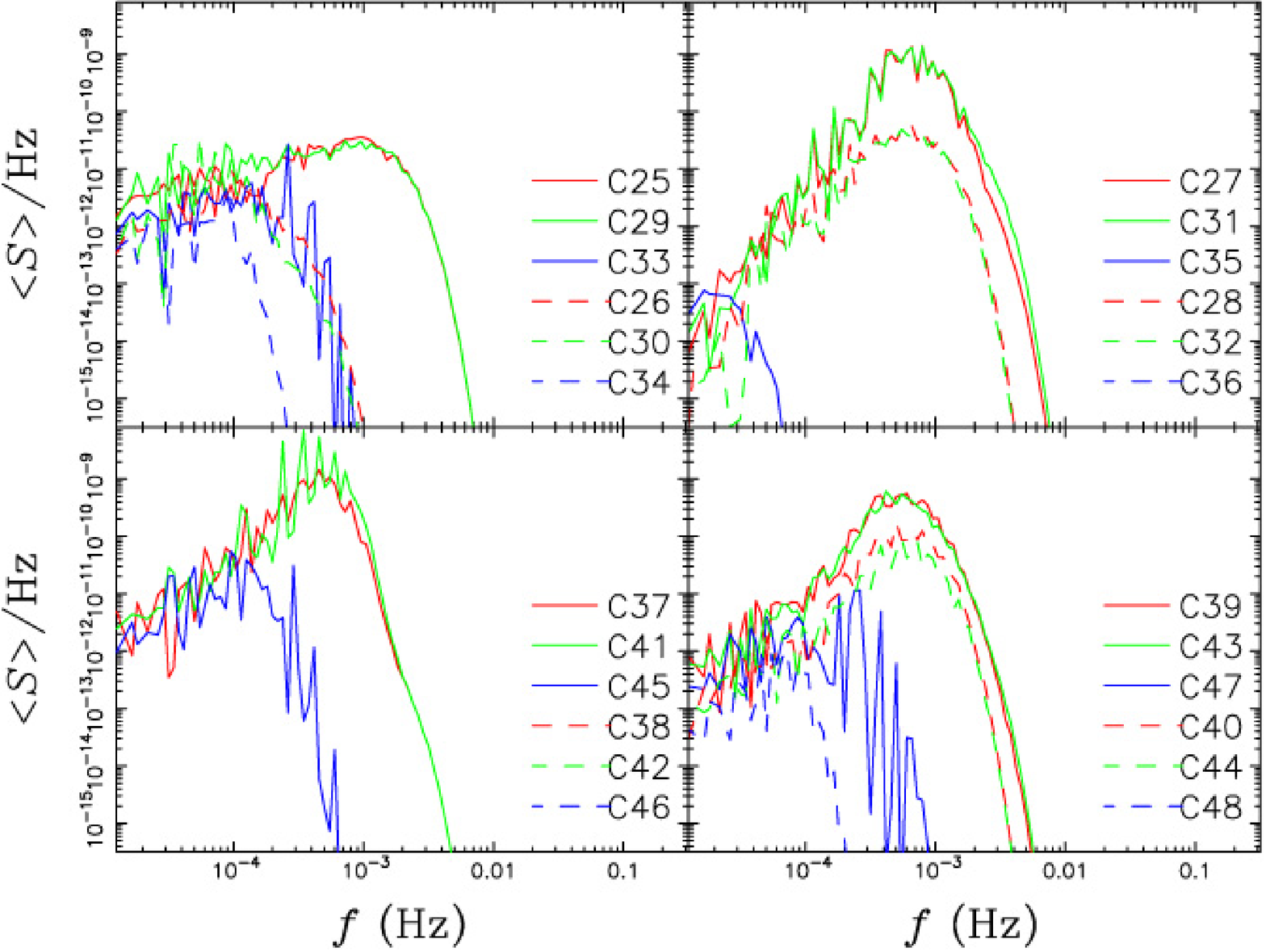}
\end{minipage}
}
\caption{The spectra distribution of gravitational wave (GW) flux calculated
by Eq.\,\ref{eq_s} from DNS population in the Galactic disc in different cases.
Note that there is no GW signals from double neutron stars (DNS) in
cases C4, C8, C10, C12 and C24, and the GW signals in cases C11, C24, C36, C38, C42 and C46
are out of the scale of the coordinates of this figure.
We take the bin size of $\Delta \log (f/{\rm Hz})=$0.03.
}
\label{fig_gwflux}
\end{figure*}

Since several authors plot gravitational energy flux rather than strain amplitude, Fig.\,\ref{fig_gwflux}
shows the same information as Fig.\,\ref{fig_gwsignal}  for DNS in terms of the
spectral energy distrubution $\langle S \rangle$, calculated from
Eq.\,\ref{eq_s}  and reduced  using the method in \S\,\ref{sec_reduction}.

\subsection{Discrete Gravitational-Wave Signals}
\label{sec_events}

In previous work \citep{Yu10,Yu11} we have used the term `resolved' GW systems to refer to those binaries
in which all of the GW strain measured within a single frequency interval over a given integration period
arises from a single system.
Since all of the DNSs  in our simulation are in eccentric orbits, their GW emission may be spread over many frequency intervals.
Thus the concept of a `resolved' DNS may not be useful.  We prefer to use the term `discrete GW signal' to
represent a frequency interval that contains a  GW signal greater than some noise threshold
over a given observation period. The {\it number} of such signals serves as a proxy to estimate
what might be expected from  future observations.

We hence define the number $\widetilde{N}_{1\sigma,3\sigma,5\sigma}$ of discrete GW signals exceeding the
noise threshold, where the subscript defines the confidence level. Thus $1\sigma$ means that  $\widetilde{N}_{1\sigma}$
frequency intervals show a GW signal with S/N$\geqslant$1, $\widetilde{N}_{3\sigma}$ gives the number of frequency intervals with
 S/N$\geqslant$3, and so on.
The results are listed in Table\,\ref{tab_ngw}. We find that a lower metallicity
and a larger IMF power-law index give a higher probability for the detection of a GW signal.
The highest probability occurs in case C15 ($Z=0.001$, $\alpha=0.5$, $\sigma=-1.5$ and constant SF
rate). This case gives 57195 discrete signals at $1\sigma$  and 19933 discrete signals at $5\sigma$.
The detection-probability for a discrete GW signal is much lower in a $\gamma-$mechanism model than
in an equivalent $\alpha-$mechanism model. For $\gamma$-mechanism models, the highest
probability occurs for C31 ($Z=0.02$, $\gamma=1.5$, $\sigma=-1.5$ and exponential SF
rate), with 75 signals at $1\sigma$ and none at $5\sigma$.

\begin{figure*}
\centering
\includegraphics[width=16cm,clip,angle=0]{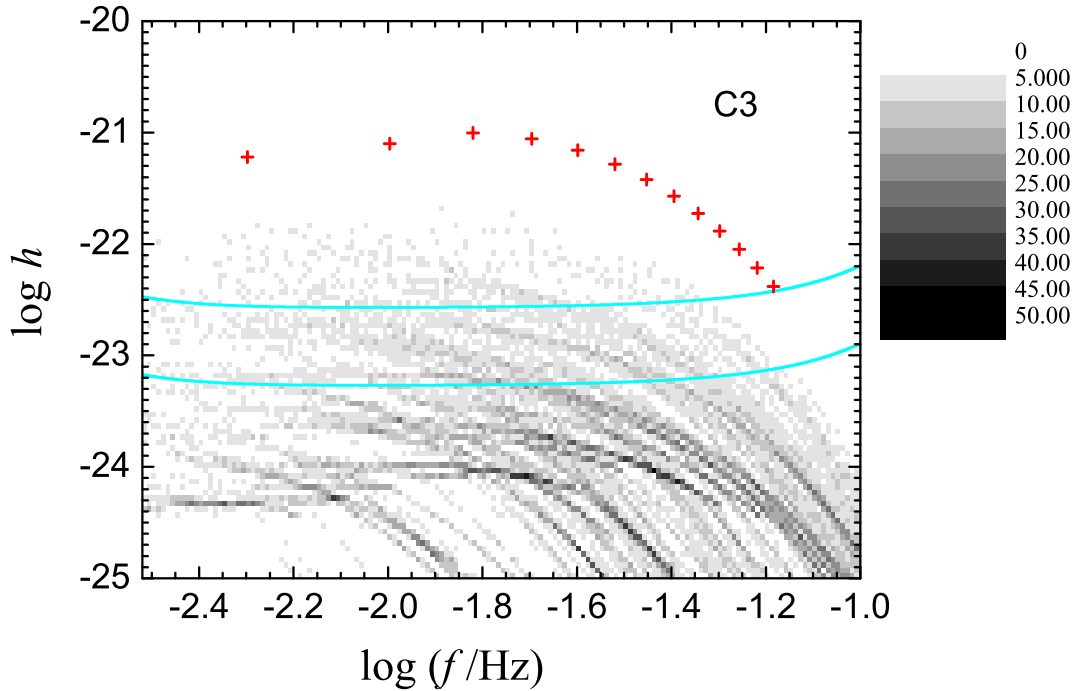}
\caption{Density map representing the gravitational wave (GW) strain amplitude of
individual double neutron stars as a function of GW frequency in case C3. The red symbols represent
the harmonic GW strain from a DNS ($m = 1.33+1.31 M_{\odot}$,
$P_{\rm orb}=198.4$ s, $e=0.508$, and $R_{\rm b}=2.81$  kpc) with S/N$>5$
({\it}i.e. at $5\sigma$ level, see section \ref{sec_events}).
The colored lines show the sensitivity of eLISA, upper line: S/N$=5$, lower line: S/N$=1$.
We adopt a  bin size $\Delta \log (f/{\rm Hz})=0.01$ and $\Delta \log h=0.05$. The grey-scale (right) shows
the numbers of GW signals per bin.}
\label{fig_hexample}
\end{figure*}

\begin{figure*}
\centering
\includegraphics[width=16cm,clip,angle=0]{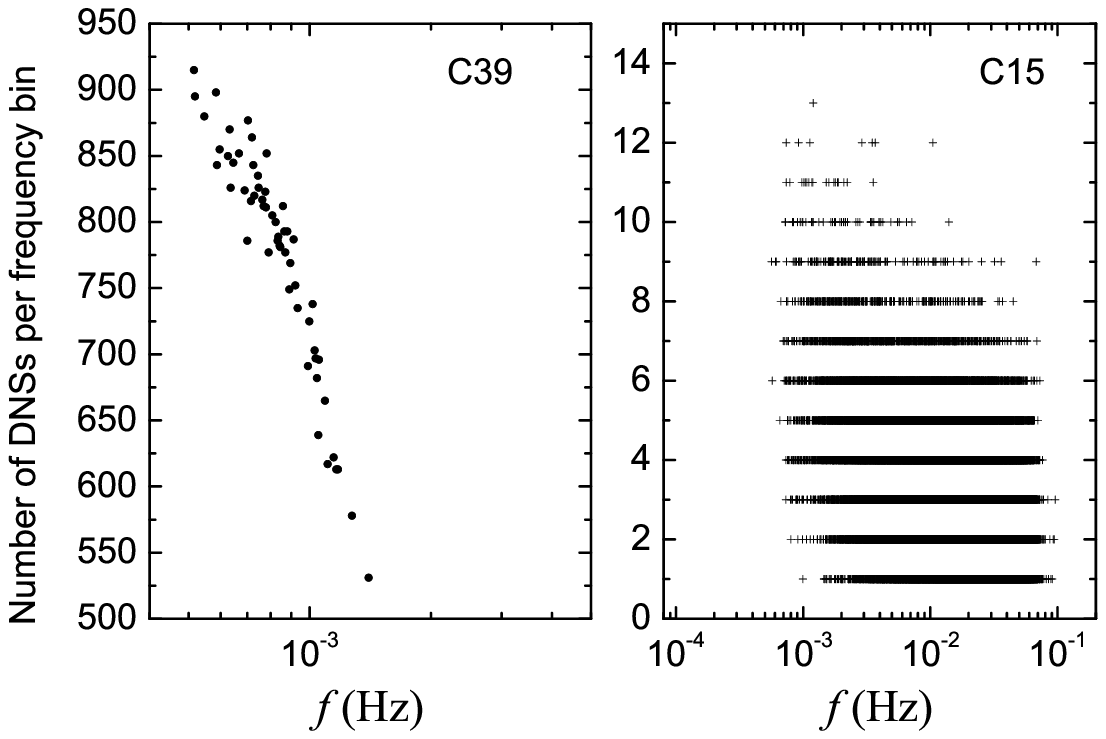}
\caption{Number of double neutron stars (DNSs) per frequency bin
as a function of gravitational wave frequency in cases C15 and C39. Note
that there may be overlap of DNSs in different frequency bins. There are 56
data points in the left panel and 57195 data points in the right panel,
corresponding to 56 and 57195 GW signals at $1\sigma$ level in Table \ref{tab_ngw}
respectively. We adopt a bin size $\Delta f ={\rm yr}^{-1}$ Hz.}
\label{fig_numexample}
\end{figure*}

Some DNSs with large eccentricities may generate a detectable GW signal at multiple
harmonics of the orbital frequency.
For example, in simulation C3, a DNS with $m = 1.33+1.31 M_{\odot}$,
$P_{\rm orb}=198.4$ s, $e=0.508$ and $R_{\rm b}=2.81$ kpc generates GW signals at the fundamental frequency
0.00504 Hz and its harmonics $n=2-13$ ({\it i.e.} 0.01008 Hz, 0.01512 Hz, 0.02016 Hz, ...) at $S/N>5$.
 The numerical error in the
frequencies is less than 0.02\%.
The detection of GW signals in harmonic series would help to constrain the DNS orbital parameters.
Note that in addition to the signal caused by the DNS in this example, there is another  weak
GW signal at frequencies of 0.02016, 0.02520, 0.04032, 0.0504 and 0.06552 Hz ($n=$4, 5, 8, 10 and 13)
respectively caused by other DNSs.

In order to investigate the potential number of `resolvable' DNSs,  Table\,\ref{tab_nn} shows
the numbers ($N$) of DNSs which have {\it at least} one detectable GW harmonic. This number is
{\it not equivalent} to the number which contribute to the GW signals in Table\,\ref{tab_ngw},
where a GW signal may exceed the detection criterion due to the combined strain of two or more DNS at the same frequency.
Table\,\ref{tab_nn} does not include DNS which generate GW signals below the detector threshold.
Hence, for example, cases C15 and C39 show 57195 and 56 GW signals (at $1\sigma$), representing
the combined contributions from about 500 - 950 and 1 - 13 DNSs in each case. In case C39, no individual DNS
generates a detectable GW signal. Figure\,\ref{fig_numexample} illustrates the number of DNSs per frequency bin
as a function of GW frequency in cases C15 and C39. We finally emphasize that the GW signals in
Table \ref{tab_ngw} come from a much smaller number of individual sources as indicated by the numbers
of resolved sources shown in Table \ref{tab_nn}. 

\begin{table}
\begin{center}
\caption{The number of gravitational wave signals in each simulation.} \label{tab_ngw}
 \begin{tabular}{llllllll}
\hline
  Case          & $\widetilde{N}_{1\sigma}$     & $\widetilde{N}_{3\sigma}$  &  $\widetilde{N}_{5\sigma}$   &    Case  &  $\widetilde{N}_{1\sigma}$  &  $\widetilde{N}_{3\sigma}$   &   $\widetilde{N}_{5\sigma}$      \\
\hline
$\alpha$ \\
  C1&    24467       &  11766           &   7253      &                   C13       &   37270       & 19660         & 11770   \\
  C2&    1127        &  625             &   399       &                   C14       &   29129       & 12777         & 6654   \\
  C3&    16406       &  8789            &   5869      &                   C15       &   57195       & 30689         & 19933   \\
  C4&    0           &  0               &   0         &                   C16       &   1622        & 759           & 476   \\
  C5&    23518       &  11454           &   7133      &                   C17       &   34057       & 16982         & 9692   \\
  C6&    870         &  482             &   306       &                   C18       &   23131       & 9397          & 4632   \\
  C7&    16371       &  8435            &   5643      &                   C19       &   54618       & 29213         & 19249   \\
  C8&    0           &  0               &   0         &                   C20       &   1492        & 693           & 449   \\
  C9&    0           &  0               &   0         &                   C21       &   0         & 0          & 0   \\
  C10&    0          &  0               &   0         &                   C22       &   0         & 0          & 0   \\
  C11&    0          &  0               &   0         &                   C23       &   0         & 0          & 0   \\
  C12&    0          &  0               &   0         &                   C24       &   0         & 0          & 0   \\
\hline
 $\gamma$ \\
  C25&    0        &  0            &   0      &                   C37       &   24       & 0          & 0   \\
  C26&    0        &  0            &   0      &                   C38       &   2        & 0          & 0   \\
  C27&    60       &  3            &   0      &                   C39       &   56       & 0          & 0   \\
  C28&    0        &  0            &   0      &                   C40       &   0        & 0          & 0   \\
  C29&    0        &  0            &   0      &                   C41       &   14       & 4          & 0   \\
  C30&    0        &  0            &   0      &                   C42       &   2        & 0          & 0   \\
  C31&    75       &  3            &   0      &                   C43       &   26       & 0          & 0   \\
  C32&    0        &  0            &   0      &                   C44       &   0        & 0          & 0   \\
  C33&    0        &  0            &   0      &                   C45       &   0        & 0          & 0   \\
  C34&    0        &  0            &   0      &                   C46       &   0        & 0          & 0   \\
  C35&    0        &  0            &   0      &                   C47       &   0        & 0          & 0   \\
  C36&    0        &  0            &   0      &                   C48       &   0        & 0          & 0   \\
\hline
\end{tabular}
\end{center}
\end{table}

\begin{table}
\begin{center}
\caption{The number of DNSs which have at least one detectable GW signal in each simulation.}
\label{tab_nn}
 \begin{tabular}{llllllll}
\hline
  Case          & ${N}_{1\sigma}$     & ${N}_{3\sigma}$  &  ${N}_{5\sigma}$   &       Case  &  ${N}_{1\sigma}$  &  ${N}_{3\sigma}$   &   ${N}_{5\sigma}$      \\
\hline
$\alpha$ \\
  C1&    1273        &  927           &   743      &                   C13       &   2434       & 1446         & 1031    \\
  C2&    45          &  42            &   35       &                   C14       &   1471       & 812          & 546   \\
  C3&    543         &  487           &   429      &                   C15       &   1473       & 1291         & 1115   \\
  C4&    0           &  0             &   0        &                   C16       &   38         & 33           & 28   \\
  C5&    1220        &  896           &   729      &                   C17       &   2005       & 1194         & 835   \\
  C6&    34          &  32            &   26       &                   C18       &   1049       &  585         & 379   \\
  C7&    538         &  474           &   409      &                   C19       &   1412       & 1226         & 1078   \\
  C8&    0           &  0             &   0        &                   C20       &   35         & 30           & 26   \\
  C9&    0           &  0             &   0        &                   C21       &   0          & 0            & 0   \\
  C10&    0          &  0             &   0        &                   C22       &   0          & 0            & 0   \\
  C11&    0          &  0             &   0        &                   C23       &   0          & 0            & 0   \\
  C12&    0          &  0             &   0        &                   C24       &   0          & 0            & 0   \\
\hline
 $\gamma$ \\
  C25&    0        &  0            &   0      &                   C37       &   2        & 0          & 0   \\
  C26&    0        &  0            &   0      &                   C38       &   1        & 0          & 0   \\
  C27&    7        &  1            &   0      &                   C39       &   3        & 0          & 0   \\
  C28&    0        &  0            &   0      &                   C40       &   0        & 0          & 0   \\
  C29&    0        &  0            &   0      &                   C41       &   2        & 1          & 0   \\
  C30&    0        &  0            &   0      &                   C42       &   2        & 0          & 0   \\
  C31&    11       &  1            &   0      &                   C43       &   1        & 0          & 0   \\
  C32&    0        &  0            &   0      &                   C44       &   0        & 0          & 0   \\
  C33&    0        &  0            &   0      &                   C45       &   0        & 0          & 0   \\
  C34&    0        &  0            &   0      &                   C46       &   0        & 0          & 0   \\
  C35&    0        &  0            &   0      &                   C47       &   0        & 0          & 0   \\
  C36&    0        &  0            &   0      &                   C48       &   0        & 0          & 0   \\
\hline
\end{tabular}
\end{center}
\end{table}

\subsection{GW signal from the Galactic components}
\label{sec_rgwgalaxy}

Amongst our experiments, cases C2, C4, C6, C8, C26, C28, C30 and C32 better represent the bulge and thin disc
for SF history, IMF and metallicity.
However, due to the lower total mass of the bulge
$\approx1-2\times10^{10}$ $M_{\odot}$ \citep{Robin03,Belczynski10b,Yu10},
the number of GW signals from the DNSs in the bulge
should be less than that in the thin disc by a factor of $2.5-5$.
Since the thick disc has a much smaller
total stellar mass ($2.6\times10^{9}$ $M_{\odot}$ \citep{Robin03,Yu10}),
the GW signal from DNSs in the thick disk
should be less than that in the thin disc by a factor of about 20.
The halo has a quite different SF history (SF only  at early epochs),
IMF ($\sigma=-1.5$), and metallicity ($Z=0.001$) from the thin disc,
but  would have a similar stellar mass ($5\times10^{10}$ $M_{\odot}$).
Numerical experiments C21, C23, C45 and C47 better reflect conditions in
 the halo. Our results show that very rare GW signals from the DNSs in the
halo can be observed, which support the results of \citet{Belczynski10b}.
Note that the influence of distance may be negligible.

In the model of \citet{Yu10}, the mean distances and standard deviations
of stars in the bulge, the disc, and the halo are, respectively, $8.50\pm0.28$ kpc,
$8.92\pm2.47$ kpc, and $8.62\pm1.13$ kpc. In fact, a realistic GW signal will lie
somewhere between the  cases studied. For example, if the mean metallicity of the thin disc lies between
0.001 and 0.02 \citep{Panter08,Belczynski10b}, a more realistic case would be a combination of
C2 and C14, or C4 and C16 and so on. We here approximate the GW contribution from the Galaxy as a whole
by correcting the thin disc contribution ($\widetilde{N}$) for the bulge ($+\widetilde{N}/2.5$),
thick disk  ($+\widetilde{N}/20$), and halo ($=0$).

\citet{Belczynski10b} concluded that the number of resolved DNSs in their population synthesis
model should be about 6, which is consistent with the results of \citet{Nelemans01b}.
For an eLISA-type observatory, and assuming case C2 (Table 4) is approximately
representative of the Galaxy as a whole,
one year of observation will yield about 1633 observable DNS-induced GW signals, 906 signals at S/N=3,
and 577 at S/N=5.

\section{Discussion}
\label{sec_discussion}

We have used an evolutionary model to simulate the birth and merger rates and the total number of DNSs
in the Galactic thin disc. Our results on rates and total number are roughly consistent with the observational
constraints \citep{Kalogera04} and with binary-star formation and evolution theory calculated by others
\citep{Nelemans01b,Dominik12}. However, \citet{Nelemans01b} only give an optimistic value for
the rates and number assuming the $\gamma-$mechanism.
\citet{Kalogera04} report results based on observation and statistical theory; they neglect
the dependency of the rates and number on initial conditions and stellar evolution parameters.
\citet{Dominik12} investigated the influence of metallicity and the $\alpha-$mechanism CE ejection coefficient,
they did not investigate different SF and IMF models.
In this paper we have  systematically investigated the influence of all these initial conditions
and stellar evolution parameters on the rates and total number of DNSs.

Using the binary-star population synthesis model, we have calculated the stable GW emission from the
long-lived DNS population which may be observable by (for example) eLISA in the frequency range
$0.001-0.1$ Hz. In fact, these calculations can be used to estimate the GW signal from other Galactic components,
{\it i.e.} the bulge, the thick disc, and the halo. The SF rates in the bulge and thin disc are most likely to be
continuous, and both share a common metallicity ($Z=0.02$). The power-law indices of the IMF
in these regions are also similar, being $-2.35$ in the bulge and $-2.5$ in the thin disc
for stars with mass $m\gtrsim1$ $M_{\odot}$ \citep{Zoccali00,Kroupa93,Kroupa01,Robin03}.

\citet{Rosado11} adopts an analytic approach to study the GW background from binary systems,
and concludes that no background signal generated by DNSs and DWDs is detectable in the frequency band
of ground-based GW detectors. Our calculations support this conclusion.

Since the GW signal from DNSs is significantly affected by SF history,
recent SF regions in the bulge and spiral arms of the Galaxy may contain many GW sources.
By comparison, although the Large and Small Magellanic Clouds may have experienced
bursts of star formation peaking roughly 2, 0.5, 0.1 and 0.012 Gyr ago \citep{Harris09} and 2.5, 0.4, and
0.06 Gyr ago \citep{Harris04}, respectively, the low Magellanic-Cloud SF rate of roughly $0.2~M_{\odot}{\rm yr}^{-1}$
implies the GW signal from Magellanic-Cloud DNSs should be negligible compared with that in the Galaxy.

A  {\it caveat} affecting all of our simulations is that, in practice, the number of detectable
DNS present in the Galaxy at any one time is small. Even in cases where the number of GW signals exceeds
$10^3$, Table \ref{tab_nn} shows that these arise from a relatively small number of `resolvable' DNS,
with eccentric systems producing signals at multiple harmonics. The DNS eccentricity distribution in our model
is essentially that defined by the common-envelope interaction physics, followed by tidal evolution as discussed
in \S 2.1.2. While observations show that highly eccentric DNS do exist at long period (cf. Table 1), tidal evolution
to short-period systems is not tested. Detailed models for the influence of kick velocity
distribution and galactic structure on the orbital parameters of DNSs and therefore their GW signal
are also needed to extend our model to different types of galaxy. More seriously, since the predicted
GW signal is based on small numbers amongst which the eccentricity distribution is determined by a Monte-Carlo
distribution applied to the parent population, uncertainties on numbers in Table 5 are more likely to scale as
$\sqrt{N}/N$, the numbers of DNS in Table 6, than as $\sqrt{\widetilde{N}}/\widetilde{N}$.

\section{Conclusion}
\label{sec_conclusion}

Using the methods of binary-star population synthesis,
we have investigated the Galactic double neutron-star (DNS) population as a function of
initial mass function, star-formation history, metallicity and common-envelope physics.
The parameter space explored is larger than covered previously, and includes
a volume corresponding to best estimates of the present-day Galaxy.
We have computed the gravitational-wave (GW) signal that would be generated by these
theoretical populations, including the multi-frequency signal generated
by DNSs in elliptical orbits. We have also explored the probability of likelihood
of detecting GW from DNS from a conceptual space-borne GW observatory (eLISA).
The main conclusions are:

(1) Observable GW from double neutron stars are more likely in low metallicity
environments, in environments where there is a high proportion of massive stars, and in regions of
recent star formation.
The first two statements are relatively obvious; low metallicity produces less luminous and, hence, smaller giants, so that binaries
are more tightly bound at the point of common-envelope ejection, whilst a higher IMF power-law index
produces more high mass stars, and therefore more DNS.
If the galactic metallicity is reduced from $Z = 0.02$ to $0.001$, the number
of observable GW signals from DNS increases by a factor up to $\approx30$ in the two continuous star
formation models ({\it e.g.} C6 and C18), and consistent with \citet{Belczynski10}. The influence of the IMF is even
stronger than that of metallicity; increasing the power-law index $\sigma$ from $-2.5$ to $-1.5$
increases the number of observable GW signals from DNS from 0 to 2434--57195 1-$\sigma$ detections
in the frequency range $0.0001 -  0.1$ Hz.
The fraction of DNS arising from star formation within the last Gyr is almost
100\% of the total DNS population.

(2) Observable GW from DNS reflect the physics of the common-envelope ejection mechanism. If
conservation of energy dominates ($\alpha-$formalism), DNS GW are more likely to be observed than if
conservation of angular momentum ($\gamma-$formalism) dominates the physics. The peak frequency at which
the average strain amplitude has a maximum value in these two mechanisms is also different. The peak
frequency for the $\alpha-$formalism is in the range of 0.001 -- 0.01 Hz, while the peak frequency for the
$\gamma-$formalism is in the range of $10^{-5} - 10^{-3}$ Hz. Observation of the DNS GW spectrum will
therefore help to constrain common-envelope ejection physics.

(3) Young DNSs most likely have eccentric orbits resulting mostly from kick velocities imparted during supernova collapse
and partly binary evolution. This creates a harmonic structure in the GW radiation of DNSs.

(4) Current observations indicate the most realistic values for the
physical parameters in the Galactic disc correspond with our cases C2 and C6, {\it i.e.} $Z=0.02,
\alpha\lambda=1.0, \sigma=-2.5$ and either constant (C2) or exponentially decreasing (C6) star formation.
We therefore expect that one year of observation with a GW observatory such as eLISA will 
detect approximately 0$-$1600 observable GW signals caused by DNSs at S/N$\geqslant$1, 0$-$900 signals at S/N$\geqslant$3,
and 0$-$570 signals at S/N$\geqslant$5 in the Galaxy between 10$^{-5}$ and 1 Hz,
coming from about 0$-$65, 0$-$60 and 0$-$50 resolved DNS.

\section*{Acknowledgments}
This work was supported by the National Science Foundation of China (NSFC),
Grant No. 11303054 and 11261140641, the Chinese Ministry of Science and Technology
under the State Key Development Program for Basic Research, Grant No. 2013CB837900,
the Projects of International Cooperation and Exchange, the key research program of
the Chinese Academy of Sciences (CAS), Grant No. KJZD-EW-T01. This work was also
supported by the Open Project Program of the key Laboratory of Radio Astronomy,
CAS and Scientific Research Foundation of the Ministry of Education.
Research at the Armagh Observatory is grant-aided by the N. Ireland
Department of Culture, Arts and Leisure.

\bibliographystyle{mn}
\bibliography{gwnsns2}

\label{lastpage}

\end{document}